\documentclass[twocolumn,prd,amssymb,aps,nofootinbib,showpacs]{revtex4}
\usepackage[usenames,dvipsnames]{color}
\usepackage{ulem}
\usepackage{graphicx}
\usepackage{amssymb}
\usepackage{amsmath}
\usepackage{epstopdf}

\def\cb{\color{black}}

\begin{document}
\title{Extracting equation of state parameters from black hole-neutron star mergers.  I.  Nonspinning black holes}
\author{Benjamin D. Lackey$^1$, Koutarou Kyutoku$^2$, Masaru Shibata$^2$, Patrick R. Brady$^1$, John L. Friedman$^1$}

\affiliation{
$^1$Department of Physics, University of Wisconsin--Milwaukee, P.O. Box 413, Milwaukee, WI 53201, USA\\
$^2$Yukawa Institute for Theoretical Physics, Kyoto University, Kyoto 606-8502, Japan
}

\begin{abstract}
The late inspiral, merger, and ringdown of a black hole-neutron star (BHNS) system can provide information 
about the neutron-star equation of state (EOS).  Candidate EOSs can be approximated by 
a parametrized piecewise-polytropic EOS above nuclear density, matched to a fixed low-density 
EOS; and we report results from a large set of 
BHNS inspiral simulations that systematically vary two parameters.  
To within the accuracy of the simulations, we find that, apart from the neutron-star mass, a 
single physical parameter $\Lambda$, describing its deformability, can be extracted from the 
late inspiral, merger, and ringdown waveform.  This parameter is related to the radius, mass, 
and $\ell=2$ Love number, $k_2$,  of the neutron star by $\Lambda = 2k_2 R^5/3M_{\rm NS}^5$, 
and it is the same parameter that determines the departure from point-particle dynamics during the 
early inspiral.  Observations of gravitational waves from BHNS inspiral thus restrict the 
EOS to a surface of constant $\Lambda$ in the parameter space, thickened by the measurement 
error.  Using various configurations of a single Advanced LIGO detector, we find that neutron stars are 
distinguishable from black holes of the same mass and that $\Lambda^{1/5}$ or equivalently $R$ can be extracted to 10--40\% accuracy from single events for mass ratios of $Q=2$ and 3 at a distance of 100~Mpc, while with the proposed 
Einstein Telescope, EOS parameters can be extracted to accuracy an order of magnitude better.
\end{abstract}


\maketitle

\section{Introduction}

Construction of the second-generation Advanced LIGO {\cb(aLIGO)} detectors is underway, and will soon begin for Advanced VIRGO and LCGT, making it likely that gravitational waveforms from compact binaries will be observed in this decade.  Plans are also in development for the third generation Einstein Telescope (ET) detector with an order-of-magnitude increase in sensitivity over aLIGO.  Population synthesis models predict that with a single aLIGO detector binary neutron star (BNS) systems will be observed with a signal-to-noise ratio (SNR) of 8, at an event rate between 0.4 and 400 times per year and with a most likely value of 40 per year~\cite{LIGORate2010}.  Black hole--neutron star (BHNS) systems are also expected, but with a more uncertain rate of between 0.2 and 300 events per year at the same SNR and with a most likely value of 10 events per year for a canonical 1.4~$M_\odot$--10~$M_\odot$ system~\cite{LIGORate2010}.  The expected mass ratios $Q = M_{\rm BH}/M_{\rm NS}$ of BHNS systems are also highly uncertain and may range from just under 3 to more than 20~\cite{Belczynskietal2010, BelczynskiKalogeraBulik2002}.  

A major goal of the gravitational-wave (GW) program is to extract from observed waveforms the physical characteristics of their sources and, in particular, to use the waveforms of inspraling and merging BNS and BHNS systems to constrain the uncertain EOS of neutron-star matter.  During inspiral the tidal interaction between the two stars leads to a small drift in the phase of the gravitational waveform relative to a point particle system.  Specifically the tidal field $\mathcal{E}_{ij}$ of one star will induce a quadrupole moment $Q_{ij}$ in the other star given by $Q_{ij} = -\lambda \mathcal{E}_{ij}$ where $\lambda$\footnote{The tidal deformability for the $\ell$th multipole is often defined in terms of the NS radius $R$ and its dimensionless $\ell$th Love number $k_\ell$ by $\lambda_\ell = \frac{2}{(2\ell-1)!!G} k_\ell R^{2\ell+1}$.  Here we will discuss only the $\ell=2$ term so we write $\lambda:=\lambda_2$.} is an EOS dependent quantity that describes how easily the star is distorted.  A method for determining $\lambda$ for relativistic stars was found by Hinderer~\cite{Hinderer2008}; its effect on the waveform was calculated to Newtonian order (with the relativistic value of $\lambda$) by Flanagan and Hinderer~\cite{FlanaganHinderer2008} and to first post-Newtonian (PN) order by Vines, Flanagan, and Hinderer~\cite{VinesFlanagan2010, VinesFlanaganHinderer2011}.  This tidal description has also been extended to higher order multipoles~\cite{DamourNagar2009tidal, BinningtonPoisson2009}.

The detectability of EOS effects have been examined for both BNS and BHNS systems using this analytical description of the inspiral.  For BNS systems, the detectability of $\lambda$ with aLIGO was examined for polytropic EOS~\cite{FlanaganHinderer2008} as well as a range of theoretical EOS commonly found in the NS literature for aLIGO and ET~\cite{HindererLackeyLangRead2010}.  These studies considered only the waveform up to frequencies of 400--500~Hz ($\sim$30--20 GW cycles before merger for 1.4~$M_\odot$ equal-mass NSs). For this early part of inspiral, they find that the tidal deformability is detectable by aLIGO only for an unusually stiff EOS and for low neutron-star masses ($<1.2~M_\odot$).  ET on the other hand would have an order of magnitude improvement in estimating $\lambda$, allowing ET to distinguish between different classes of EOS.  For BHNS systems, using the recently calculated 1PN corrections, Pannarale et al.~\cite{PannaraleRezzollaOhmeRead2011} examined detectability for a range of mass ratios, finding that aLIGO will be able to distinguish between BHNS and binary black hole (BBH) systems only for low mass ratios and stiff EOS when considering the full inspiral waveform up to the point of tidal disruption.

In sharp contrast to these analytic post-Newtonian results, analysis of just the last few orbits of BNS inspiral from numerical simulations has shown that the NS radius may be extracted to a higher accuracy, of $\mathcal{O}(10\%)$~\cite{ReadMarkakisShibata2009}, and this is confirmed by a study based on a set of longer and more accurate waveforms from two different codes~\cite{Read2011}.  In addition, comparisons between the analytical tidal description and BNS quasiequilibrium sequences~\cite{DamourNagar2010} as well as long BNS numerical waveforms~\cite{BaiottiDamour2010, BaiottiDamour2011} suggest that corrections beyond the 1PN quadrupole description are significant and substantially increase the tidal effect during the late inspiral.

Numerical BHNS simulations have also been done to examine the dependence of the waveform on mass ratio, BH spin, NS mass, and the neutron-star EOS~\cite{ShibataUryu2006, Etienne2007, ShibataUryu2007, ShibataTaniguchi2008, Duez2008, Etienne2009, ShibataKyutokuYamamotoTaniguchi2009, ShibataKyutoku2010, KyutokuShibataTaniguchi2010, Duez2010, FoucartDuezKidderTeukolsky2011, KyutokuOkawaShibataTaniguchi2011}.  However, an analysis of the detectability of EOS information with GW detectors using these simulations has not yet been done, and the present paper presents the first results of this kind.  EOS information from tidal interactions is present in the inspiral waveform.  For BHNS systems, however, the stronger signal is likely to arise from a sharp drop in the GW amplitude arising from tidal disruption prior to merger or, when there is negligible disruption, from the cutoff frequency at merger~\cite{Vallisneri2000}.

We find from simulations of the last few orbits, merger, and ringdown of BHNS systems with varying EOS that, to within numerical accuracy, the EOS parameter extracted from the waveform is the same tidal parameter $\Lambda$ that determines the departure from point particle behavior during inspiral; here $\Lambda$ is a dimensionless version of the tidal parameter:
\begin{equation}
\Lambda := G\lambda \left(\frac{c^2}{GM_{\rm NS}}\right)^5 
= \frac{2}{3} k_2 \left(\frac{c^2R}{GM_{\rm NS}}\right)^5, 
\end{equation}
where $k_2$ is the quadrupole Love number.

The constraint on the EOS imposed by gravitational-wave observations of BHNS inspiral and merger is essentially a restriction of the space of EOS $p=p(\rho)$ 
to a hypersurface of constant $\Lambda$, thickened by the uncertainty in the measurement (that is, 
a restriction to the set of EOS for which a spherical 
neutron star of the mass observed in the inspiral has tidal parameter $\Lambda$).  
We use a parametrized EOS based on piecewise polytropes \cite{ReadLackey2009}, to delineate 
this region in the EOS space, but the result can be used to constrain any choice of parameters 
for the EOS space.

In Sec.~\ref{sec:parametrized} we discuss the parametrized EOS used in the simulations.  We give in Sec.~\ref{sec:numerical} an overview of the numerical methods used and, in Sec.~\ref{sec:waveforms}, a description of the waveforms from the simulations.  We then discuss the analytical waveforms used for the early inspiral and issues related to joining the analytical and numerical waveforms to create hybrids in Sec.~\ref{sec:hybrid}, and we then estimate the uncertainty in extracting EOS parameters in Sec.~\ref{sec:errors}.  Finally, we discuss future work in Sec.~\ref{sec:discussion}.  In the appendices we discuss methods for numerically evaluating the Fisher matrix, and we provide instructions for generating effective one body (EOB) waveforms.  In a second paper we will consider the detectability of EOS parameters for BHNS systems with spinning BHs.

{\it Conventions}:  Unless otherwise stated we set $G=c=1$.  Base 10 and base $e$ logarithms are denoted $\log$ and $\ln$ respectively.  We define the Fourier transform $\tilde{x}(f)$ of a function $x(t)$ by
\begin{equation}
\tilde x(f)=\int_{-\infty}^\infty x(t) e^{-2 \pi i f t}\,dt,
\end{equation}
and the inverse Fourier transform by
\begin{equation}
x(t)=\int_{-\infty}^\infty \tilde x(f) e^{2 \pi i f t}\,df.
\end{equation}

\section{Parametrized EOS}
\label{sec:parametrized}

To understand the dependence of the BHNS waveform on the EOS we systematically vary the free parameters of a parametrized EOS and then simulate a BHNS inspiral for each set of parameters.  We choose the piecewise polytropic EOS introduced in Ref~\cite{ReadLackey2009}.  Within each density interval $\rho_{i-1} < \rho < \rho_i$, the pressure $p$ is given in terms of the rest mass density $\rho$ by
\begin{equation}
p(\rho)=K_i \rho^{\Gamma_i},
\end{equation}
where the adiabatic index $\Gamma_i$ is constant in each interval, and the pressure constant $K_i$ is chosen so that the EOS is continuous at the boundaries $\rho_i$ between adjacent segments of the EOS.  The energy density $\epsilon$ is found using the first law of thermodynamics,
\begin{equation}
d\frac{\epsilon}{\rho} = -p d\frac{1}{\rho}.
\end{equation}

Ref.~\cite{ReadLackey2009} uses a fixed low density EOS for the NS crust.  The parametrized high density EOS is then joined onto the low density EOS at a density $\rho_0$ that depends on the values of the high-density EOS parameters.  The high-density EOS consists of a three-piece polytrope with fixed dividing densities $\rho_1=10^{14.7}$~g/cm$^3$ and $\rho_2=10^{15}$~g/cm$^3$ between the three polytropes.  The resulting EOS has four free parameters.  The first parameter, the pressure $p_1$ at the first dividing density $\rho_1$, is closely related to the radius of a 1.4~$M_\odot$ NS~\cite{LattimerPrakash2001}.  The other three parameters are the adiabatic indices $\{\Gamma_1, \Gamma_2, \Gamma_3\}$ for the three density intervals.  This parametrization accurately fits a wide range of theoretical EOS and reproduces the corresponding NS properties such as radius, moment of inertia, and maximum mass to a few percent~\cite{ReadLackey2009}.

Following previous work on BNS~\cite{ReadMarkakisShibata2009} and BHNS simulations~\cite{KyutokuShibataTaniguchi2010, KyutokuOkawaShibataTaniguchi2011} we use a simplified two-parameter version of the piecewise-polytrope parametrization and uniformly vary each of these parameters.  For our two parameters we use the pressure $p_1$ as well as a single fixed adiabatic index $\Gamma = \Gamma_1 = \Gamma_2 = \Gamma_3$ for the core.  The crust EOS is given by a single polytrope with the constants $K_0 = 3.5966\times 10^{13}$ in cgs units and $\Gamma_0 = 1.3569$ so that the pressure at $10^{13}$~g/cm$^3$ is $1.5689\times 10^{31}$~dyne/cm$^2$. (For most values of $p_1$, $\Gamma_1$, and $\Gamma_2$, the central density of a 1.4~$M_\odot$ star is below or just above $\rho_2$, so the parameter $\Gamma_3$ is irrelevant anyway for BNS before merger and BHNS for all times.) 

\begin{figure}[!htb]
\begin{center}
\includegraphics[width=80mm]{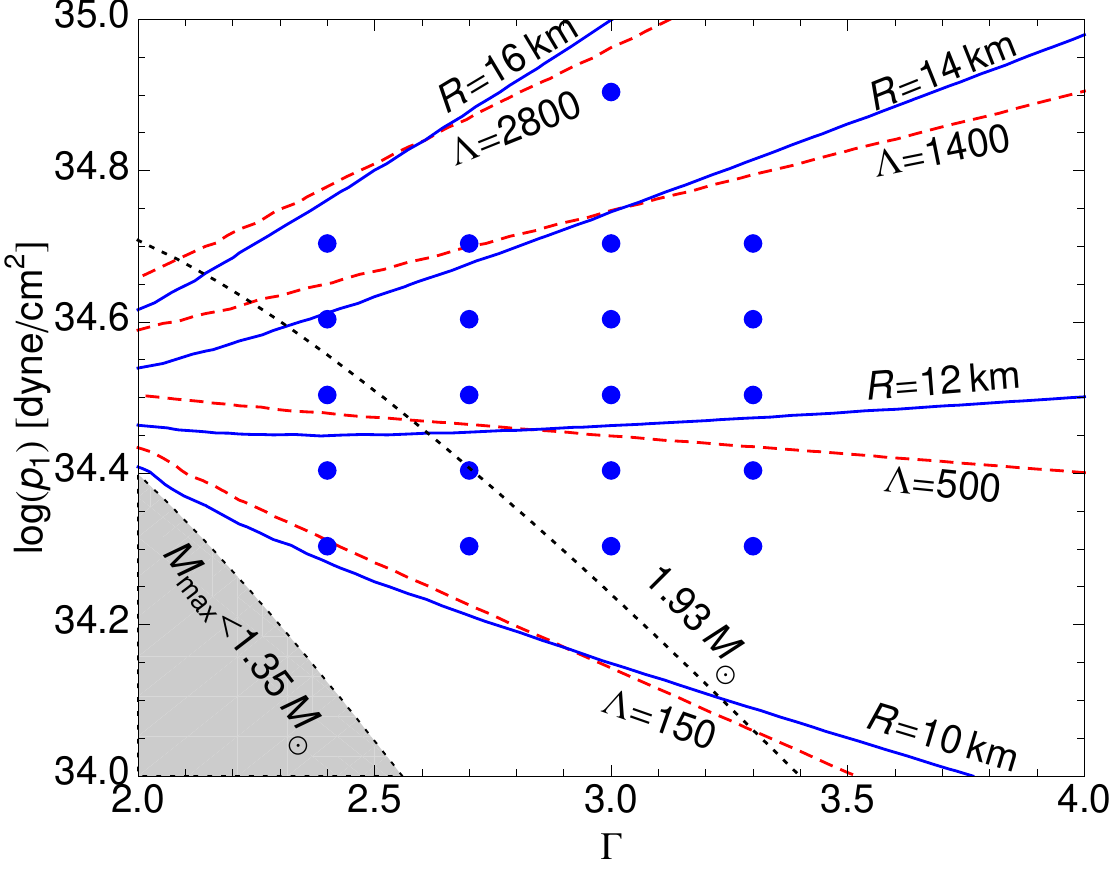}
\end{center}
\caption{ \label{fig:paramspace}
The 21 EOS used in the simulations are represented by blue points in the parameter space.  For a NS of mass 1.35~$M_\odot$, contours of constant radius are solid blue and contours of constant tidal deformability $\Lambda$ are dashed red.  Also shown are dotted contours of maximum NS mass.  The shaded region does not allow a 1.35~$M_\odot$ NS.
}
\end{figure}

We list in Table~\ref{tab:properties} the 21 EOS used in the simulations along with some of the NS properties.  In addition, we plot the EOS as points in parameter space in Fig.~\ref{fig:paramspace} along with contours of constant radius, tidal deformability $\Lambda$, and maximum NS mass.  The 1.93~$M_\odot$ maximum mass contour corresponds to the recently observed pulsar with a mass of $1.97\pm0.04~M_\odot$ measured using the Shapiro delay~\cite{DemorestPennucciRansom2010}.  In this two-parameter cross section of the full four-parameter EOS space, parameters below this curve are ruled out.

\begin{table*}[!htb]
\caption{ \label{tab:properties}
Neutron star properties for the 21 EOS used in the simulations.  The original EOS names~\cite{ReadMarkakisShibata2009, KyutokuShibataTaniguchi2010, KyutokuOkawaShibataTaniguchi2011} are also listed.  $p_1$ is given in units of~dyne/cm$^2$, maximum mass is in $M_\odot$, and neutron star radius $R$ is in~km.  $R$, $k_2$, and $\Lambda$ are given for the two masses used: $\{1.20, 1.35\}~M_\odot$.  The values listed for $\log p_1$ are rounded to three digits.  The exact values used in the simulations can be found by adding $\log (c/{\rm cm\ s}^{-1})^2 - 20.95 \approx 0.00364$ (e.g.\ 34.3 becomes 34.30364).
}
\begin{center}
\begin{tabular}{llccc|ccr|ccr}
\hline\hline
\multicolumn{2}{c}{EOS} & $\log p_1$ & $\Gamma$ & $M_{\rm max}$ & $R_{1.20}$ & $k_{2, 1.20}$ & $\Lambda_{1.20}$ & $R_{1.35}$ & $k_{2, 1.35}$ & $\Lambda_{1.35}$\\
\hline
 \text{p.3$\Gamma $2.4} & Bss & 34.3 & 2.4 & 1.566 & 10.66 & 0.0765 & 401 & 10.27 & 0.0585 & 142 \\
 \text{p.3$\Gamma $2.7} & Bs & 34.3 & 2.7 & 1.799 & 10.88 & 0.0910 & 528 & 10.74 & 0.0751 & 228 \\
 \text{p.3$\Gamma $3.0} & B & 34.3 & 3.0 & 2.002 & 10.98 & 0.1010 & 614 & 10.96 & 0.0861 & 288 \\
 \text{p.3$\Gamma $3.3} &  & 34.3 & 3.3 & 2.181 & 11.04 & 0.1083 & 677 & 11.09 & 0.0941 & 334 \\
 \text{p.4$\Gamma $2.4} & HBss & 34.4 & 2.4 & 1.701 & 11.74 & 0.0886 & 755 & 11.45 & 0.0723 & 301 \\
 \text{p.4$\Gamma $2.7} & HBs & 34.4 & 2.7 & 1.925 & 11.67 & 0.1004 & 828 & 11.57 & 0.0855 & 375 \\
 \text{p.4$\Gamma $3.0} & HB & 34.4 & 3.0 & 2.122 & 11.60 & 0.1088 & 872 & 11.61 & 0.0946 & 422 \\
 \text{p.4$\Gamma $3.3} &  & 34.4 & 3.3 & 2.294 & 11.55 & 0.1151 & 903 & 11.62 & 0.1013 & 454 \\
 \text{p.5$\Gamma $2.4} &  & 34.5 & 2.4 & 1.848 & 12.88 & 0.1000 & 1353 & 12.64 & 0.0850 & 582 \\
 \text{p.5$\Gamma $2.7} &  & 34.5 & 2.7 & 2.061 & 12.49 & 0.1096 & 1271 & 12.42 & 0.0954 & 598 \\
 \text{p.5$\Gamma $3.0} & H & 34.5 & 3.0 & 2.249 & 12.25 & 0.1165 & 1225 & 12.27 & 0.1029 & 607 \\
 \text{p.5$\Gamma $3.3} &  & 34.5 & 3.3 & 2.413 & 12.08 & 0.1217 & 1196 & 12.17 & 0.1085 & 613 \\
 \text{p.6$\Gamma $2.4} &  & 34.6 & 2.4 & 2.007 & 14.08 & 0.1108 & 2340 & 13.89 & 0.0970 & 1061 \\
 \text{p.6$\Gamma $2.7} &  & 34.6 & 2.7 & 2.207 & 13.35 & 0.1184 & 1920 & 13.32 & 0.1051 & 932 \\
 \text{p.6$\Gamma $3.0} &  & 34.6 & 3.0 & 2.383 & 12.92 & 0.1240 & 1704 & 12.97 & 0.1110 & 862 \\
 \text{p.6$\Gamma $3.3} &  & 34.6 & 3.3 & 2.537 & 12.63 & 0.1282 & 1575 & 12.74 & 0.1155 & 819 \\
 \text{p.7$\Gamma $2.4} &  & 34.7 & 2.4 & 2.180 & 15.35 & 0.1210 & 3941 & 15.20 & 0.1083 & 1860 \\
 \text{p.7$\Gamma $2.7} &  & 34.7 & 2.7 & 2.362 & 14.26 & 0.1269 & 2859 & 14.25 & 0.1144 & 1423 \\
 \text{p.7$\Gamma $3.0} & 1.5H & 34.7 & 3.0 & 2.525 & 13.62 & 0.1313 & 2351 & 13.69 & 0.1189 & 1211 \\
 \text{p.7$\Gamma $3.3} &  & 34.7 & 3.3 & 2.669 & 13.20 & 0.1346 & 2062 & 13.32 & 0.1223 & 1087 \\
 \text{p.9$\Gamma $3.0} & 2H & 34.9 & 3.0 & 2.834 & 15.12 & 0.1453 & 4382 & 15.22 & 0.1342 & 2324 \\
\hline\hline
\end{tabular}
\end{center}
\end{table*}

\section{Numerical methods}
\label{sec:numerical}

We employ BHNS binaries in quasiequilibruim states for initial conditions of our numerical simulations. We compute a quasiequilibrium state of the BHNS binary as a solution of the initial value problem of general relativity, employing the piecewise polytopic EOS described in the previous section. The details of the formulation and numerical methods are described in Refs.~\cite{KyutokuShibataTaniguchi2009, KyutokuShibataTaniguchi2010}. Computations of the quasiequilibrium states are performed using the spectral-method library {\tt LORENE}~\cite{LORENE}.

Numerical simulations are performed using an adaptive-mesh refinement code {\tt SACRA}~\cite{YamamotoShibataTaniguchi2008}. {\tt SACRA} solves the Einstein evolution equations in the BSSN formalism with the moving puncture gauge, and solves the hydrodynamic equations with a high-resolution central scheme. The formulation, the gauge conditions, and the numerical scheme are the same as those described in Ref.~\cite{KyutokuShibataTaniguchi2010}. For the EOS, we decompose the pressure and energy density into cold and thermal parts as
\begin{equation}
 p = p_{\rm cold} + p_{\rm th} \; , \; \epsilon = \epsilon_{\rm cold} +
  \epsilon_{\rm th} .
\end{equation} 
We calculate the cold parts of both variables using the piecewise polytropic EOS from $\rho$, and then the thermal part of the energy density is defined from $\epsilon$ as $\epsilon_{\rm th} = \epsilon - \epsilon_{\rm cold}$. Because $\epsilon_{\rm th}$ vanishes in the absence of shock heating, $\epsilon_{\rm th}$ is regarded as the finite temperature part. In our simulations, we adopt a $\Gamma$-law ideal gas
EOS
\begin{equation}
 p_{\rm th} = ( \Gamma_{\rm th} - 1 ) \epsilon_{\rm th} ,
\end{equation} to determine the thermal part of the pressure, and choose $\Gamma_{\rm th}$ equal to the adiabatic index in the crust region, $\Gamma_0$, for simplicity.

In our numerical simulations, gravitational waves are extracted by calculating the outgoing part of the Weyl scalar $\Psi_4$ at finite coordinate radii $\sim 400 M_\odot$, and by integrating $\Psi_4$ twice in time as
\begin{equation}
 h_+ (t) - i h_\times (t) =  \int_{-\infty}^t dt' \int_{-\infty}^{t'} dt''\, \Psi_4 ( t'' )
  . \label{eq:wave}
\end{equation}
In this work, we perform this time integration with a ``fixed frequency integration'' method to eliminate unphysical drift components in the waveform~\cite{ReisswigPollney2010}. In this method, we first perform a Fourier transformation of $\Psi_4$ as
\begin{equation}
\tilde{\Psi}_4 ( f ) = \int_{-\infty}^\infty dt\, \Psi_4 (t) e^{-2\pi i f t} .
\end{equation}
Using this, Eq.~(\ref{eq:wave}) is rewritten as
\begin{equation}
 h_+ (t) - i h_\times (t) = - \frac{1}{(2 \pi)^2} \int_{-\infty}^\infty d f\, \frac{\tilde{\Psi}_4 (
  f )}{f^2} e^{2 \pi i f t} .
\end{equation}
We then replace $1 / f^2$ of the integrand with $1 / f_0^2$ for $|f| < f_0$, where $f_0$ is a free parameter in this method. By appropriately choosing $f_0$, this procedure suppresses unphysical, low-frequency components of gravitational waves. As proposed in Ref.~\cite{ReisswigPollney2010}, we choose $f_0$ to be $\sim 0.8 m \Omega_0/2\pi$, where $\Omega_0$ is the initial orbital angular velocity and $m(=2)$ is the azimuthal quantum number.

\section{Description of waveforms}
\label{sec:waveforms}

Using the 21 EOS described in Table~\ref{tab:properties}, we have performed 30 BHNS inspiral and merger simulations with different mass ratios $Q = M_{\rm BH}/M_{\rm NS}$ and neutron star masses $M_{\rm NS}$.  A complete list of these simulations is given in Table~\ref{tab:simulations}.  For the mass ratio $Q=2$ and NS mass $M_{\rm NS}=1.35~M_\odot$, we performed a simulation for each of the 21 EOS.  In addition, we performed simulations of a smaller NS mass ($Q=2$, $M_{\rm NS}=1.20~M_\odot$) and a larger mass ratio ($Q=3$, $M_{\rm NS}=1.35~M_\odot$), in which we only varied the pressure $p_1$ over the range $34.3 \le \log(p_1 / ({\rm dyne\ cm}^{-2})) \le 34.9$ while holding the core adiabatic index fixed at $\Gamma = 3.0$. 

\begin{table}[!htb]
\caption{ \label{tab:simulations}
Data for the 30 BHNS simulations.  NS mass is in units of $M_\odot$, and $\Omega_0M$ is the angular velocity used in the initial data where $M = M_{\rm BH} + M_{\rm NS}$.  
}
\begin{center}
\begin{tabular}{cccc|cccc}
\hline\hline
$Q$ & $M_{\rm NS}$ & EOS & $\Omega_0M$ & $Q$ & $M_{\rm NS}$ & EOS & $\Omega_0M$ \\
\hline
2 & 1.35 & p.3$\Gamma $2.4 & 0.028 & 2 & 1.35 & p.6$\Gamma $3.3 & 0.025 \\
2 & 1.35 & p.3$\Gamma $2.7 & 0.028 & 2 & 1.35 & p.7$\Gamma $2.4 & 0.025 \\
2 & 1.35 & p.3$\Gamma $3.0 & 0.028 & 2 & 1.35 & p.7$\Gamma $2.7 & 0.025 \\
2 & 1.35 & p.3$\Gamma $3.3 & 0.025 & 2 & 1.35 & p.7$\Gamma $3.0 & 0.028 \\
2 & 1.35 & p.4$\Gamma $2.4 & 0.028 & 2 & 1.35 & p.7$\Gamma $3.3 & 0.025 \\
2 & 1.35 & p.4$\Gamma $2.7 & 0.028 & 2 & 1.35 & p.9$\Gamma $3.0 & 0.025 \\ \cline{5-8}
2 & 1.35 & p.4$\Gamma $3.0 & 0.028 & 2 & 1.20 & p.3$\Gamma $3.0 & 0.028 \\
2 & 1.35 & p.4$\Gamma $3.3 & 0.025 & 2 & 1.20 & p.4$\Gamma $3.0 & 0.028 \\
2 & 1.35 & p.5$\Gamma $2.4 & 0.025 & 2 & 1.20 & p.5$\Gamma $3.0 & 0.028 \\
2 & 1.35 & p.5$\Gamma $2.7 & 0.025 & 2 & 1.20 & p.9$\Gamma $3.0 & 0.022 \\ \cline{5-8}
2 & 1.35 & p.5$\Gamma $3.0 & 0.028 & 3 & 1.35 & p.3$\Gamma $3.0 & 0.030 \\
2 & 1.35 & p.5$\Gamma $3.3 & 0.025 & 3 & 1.35 & p.4$\Gamma $3.0 & 0.030 \\
2 & 1.35 & p.6$\Gamma $2.4 & 0.025 & 3 & 1.35 & p.5$\Gamma $3.0 & 0.030 \\
2 & 1.35 & p.6$\Gamma $2.7 & 0.025 & 3 & 1.35 & p.7$\Gamma $3.0 & 0.030 \\
2 & 1.35 & p.6$\Gamma $3.0 & 0.025 & 3 & 1.35 & p.9$\Gamma $3.0 & 0.028 \\
\hline\hline
\end{tabular}
\end{center}
\end{table}

Two of the gravitational waveforms are shown in Fig.~\ref{fig:waveform} below.  The waveforms are compared with EOB BBH waveforms of the same mass ratio and NS mass which are also shown.  Specifically we use the EOB formalism discussed in Appendix~\ref{app:eob}. The most significant differences begin just before the merger of the black hole and neutron star.  For neutron stars with a small radius, the black hole does not significantly distort the neutron star which crosses the event horizon intact.  As a result, the merger and ringdown of these waveforms are very similar to the BBH waveform.  However, a larger NS may be completely tidally disrupted just before merger resulting in a supressed merger and ringdown waveform.
Disruption suppresses the ringdown for two reasons related to the spreading of the matter: 
The ringdown is primarily a superposition of nonaxisymmetric quasinormal modes, 
dominated by the $l=m=2$ mode, while the disrupted matter is roughly axisymmetric 
as it accretes onto the black hole; and the accretion timescale of the spread-out
matter is long compared to the periods of the dominant modes.  

\begin{figure*}[!htb]
\begin{center}
\includegraphics[width=160mm]{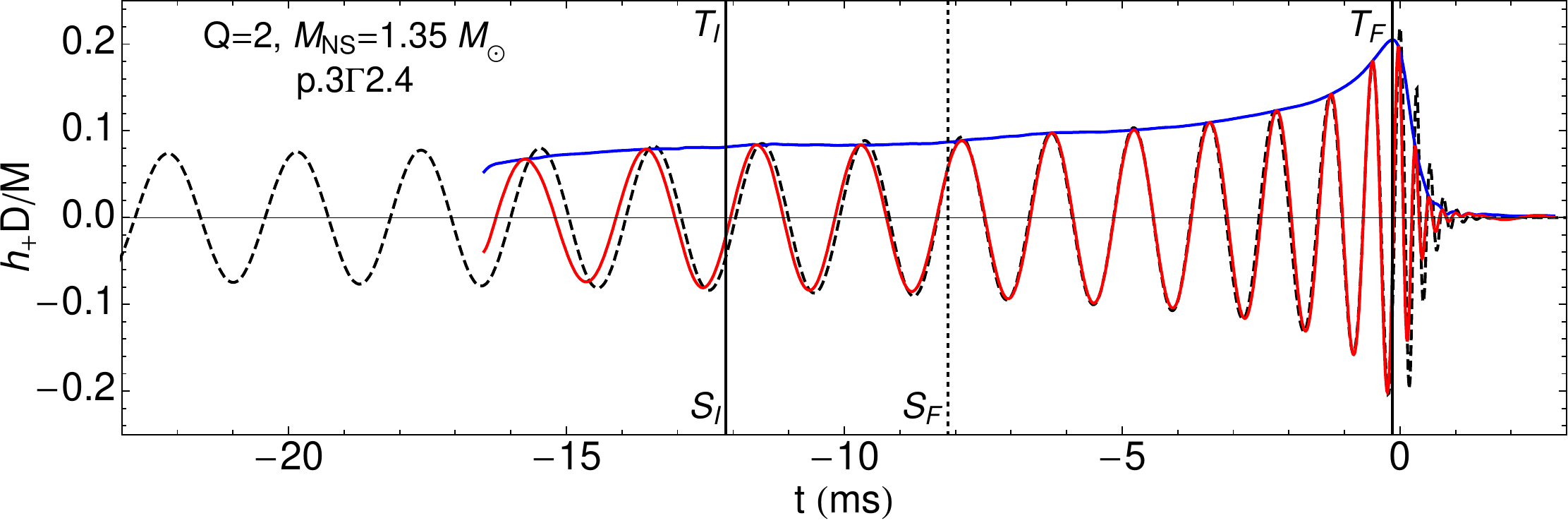}
\includegraphics[width=160mm]{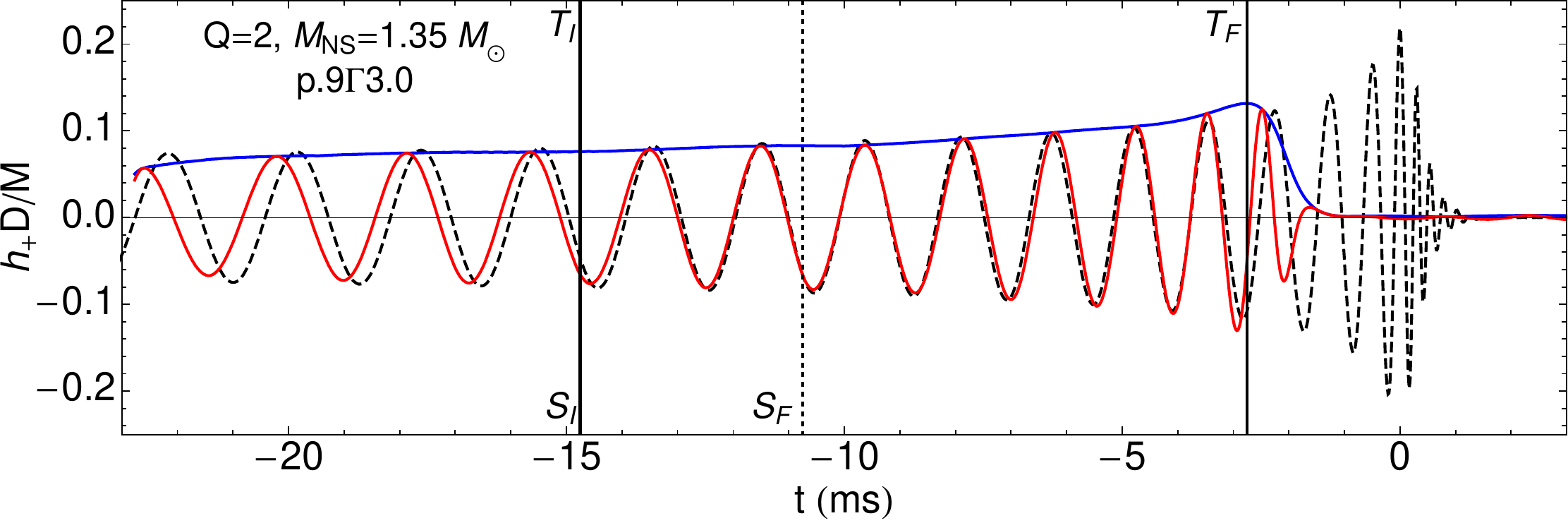}
\end{center}
\caption{ \label{fig:waveform}
$h_+$ and $|h|=|h_+-ih_\times|$ for BHNS waveforms for $(Q=2, M_{\rm NS}=1.35~M_\odot)$ with two different EOS are represented by solid red and blue curves respectively.  The softest EOS p.3$\Gamma$2.4 is on top and the stiffest EOS p.9$\Gamma$3.0 is on bottom.  An EOB BBH waveform (dashed) with the same values of $Q$ and $M_{\rm NS}$ is matched to each numerical waveform within the matching window $T_I<t<T_F$ bounded by solid vertical lines.  A hybrid EOB BBH--Numerical BHNS waveform is generated by splicing the waveforms together within a splicing window $S_I<t<S_F$ bounded by dotted vertical lines.  The matching window is 12~ms long and ends at the numerical merger time $t_M^{\rm NR}$ (time when the numerical waveform reaches its maximum amplitude), while the splicing window is 4~ms long and begins at the start of the matching window ($S_I = T_I$).
}
\end{figure*}

\begin{figure}[!htb]
\begin{center}
\includegraphics[width=80mm]{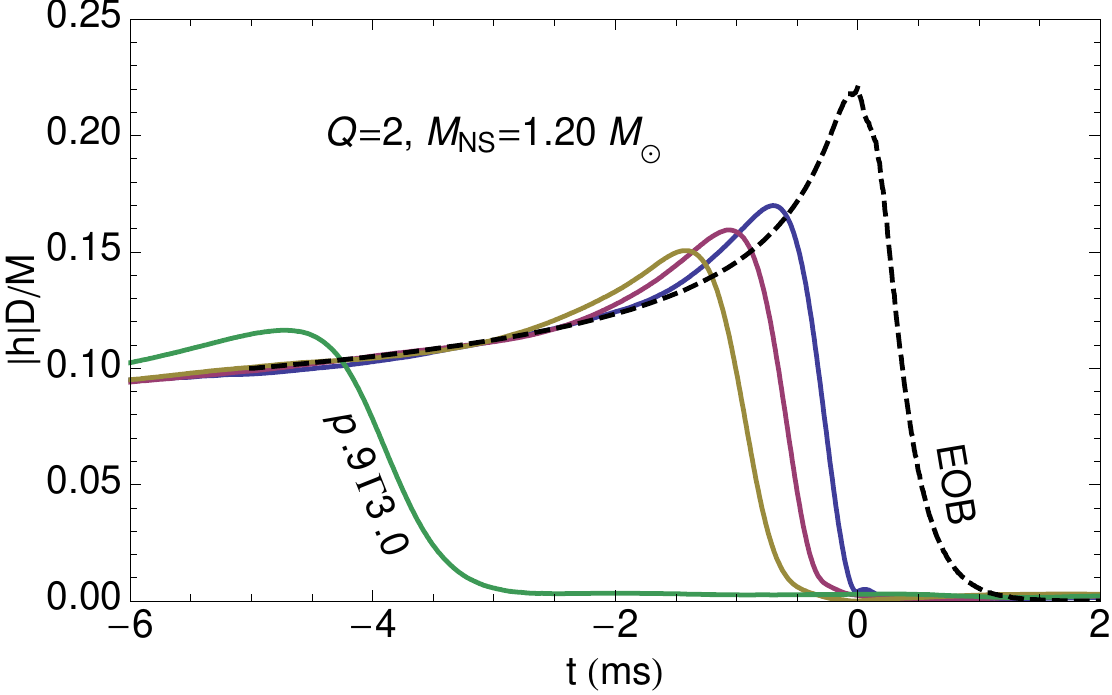}
\includegraphics[width=80mm]{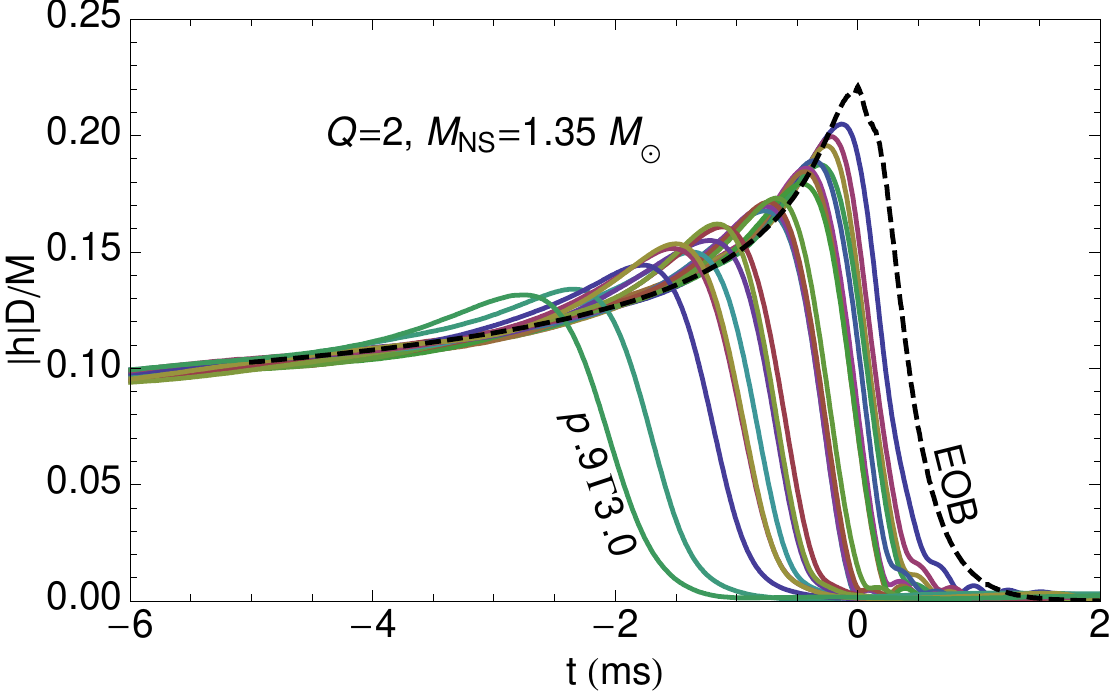}
\includegraphics[width=80mm]{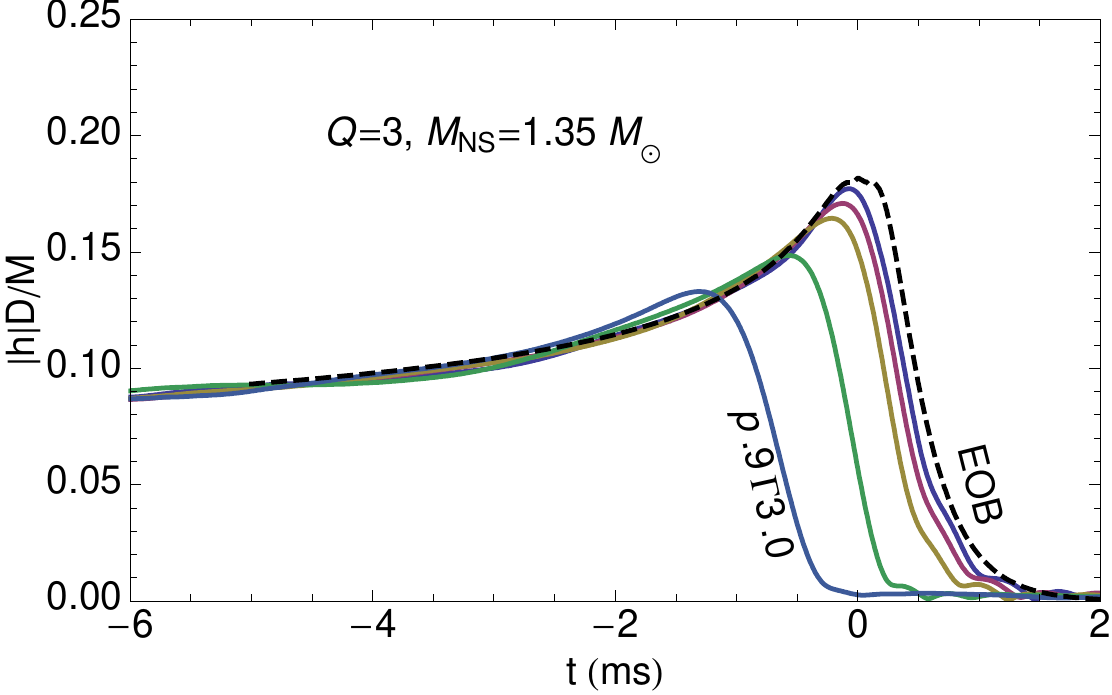}
\end{center}
\caption{ \label{fig:hamp}
Amplitude of the complex waveform $h = h_+ - ih_\times$.  Dashed curves are EOB waveforms with the same $Q$ and $M_{\rm NS}$.  Matching and splicing conventions are those of Fig.~\ref{fig:waveform}.
}
\end{figure}

\begin{figure}[!htb]
\begin{center}
\includegraphics[width=80mm]{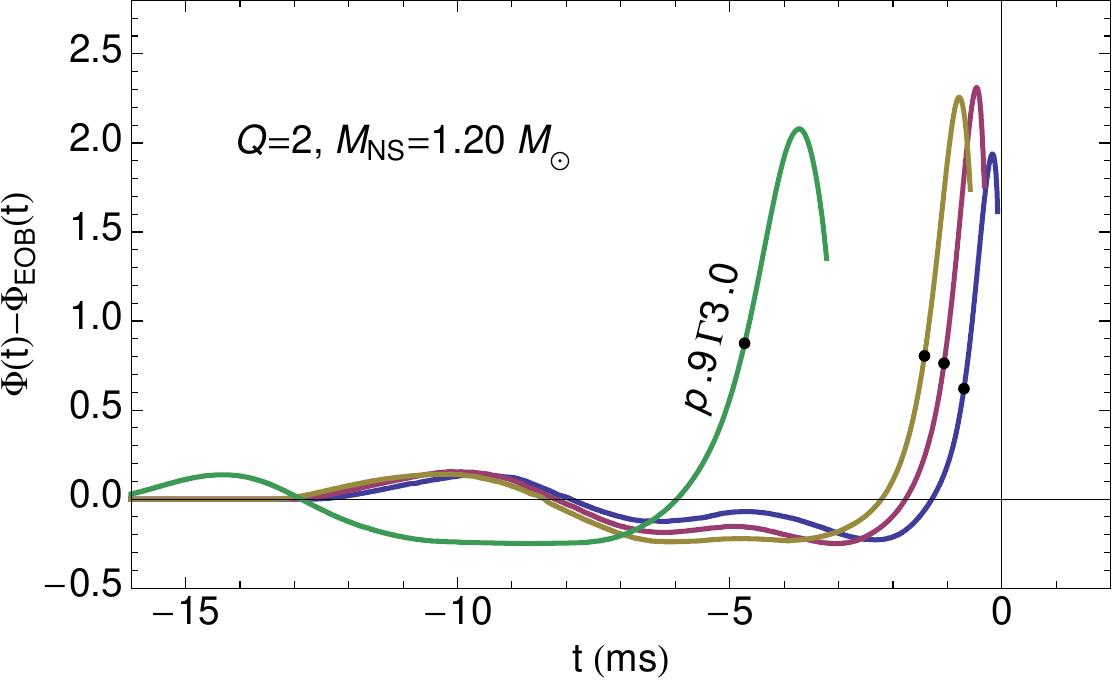}
\includegraphics[width=80mm]{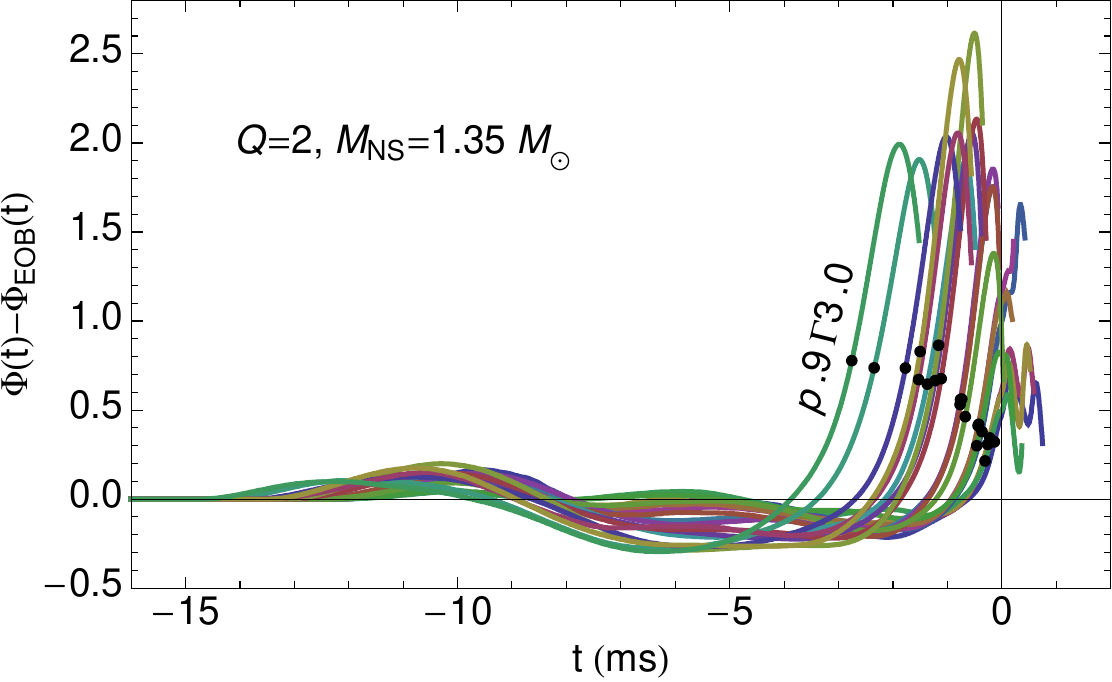}
\includegraphics[width=80mm]{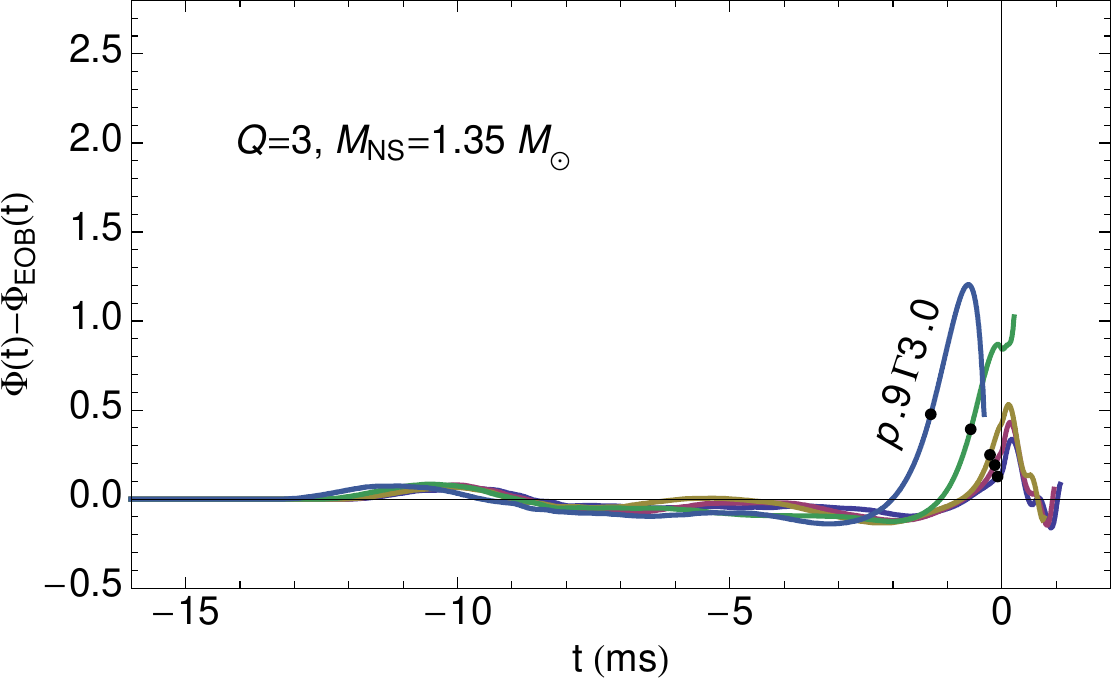}
\end{center}
\caption{ \label{fig:phasetime}
Cumulative phase difference $\Phi-\Phi_{\rm EOB}$ between BHNS waveform and EOB BBH waveform with the same $Q$ and $M_{\rm NS}$.  The phase is defined by breaking up each complex waveform into amplitude and cumulative phase $h_+(t) - i h_\times(t) = A(t)e^{-i\Phi(t)}$.  The black point on each curve indicates the BHNS merger time $t_M^{\rm NR}$ defined as the time of maximum amplitude $A(t_M^{\rm NR})$.  The curve is truncated when the amplitude $AD/M$ drops below 0.01.  Matching and splicing conventions are those of Fig.~\ref{fig:waveform}.
}
\end{figure}

The dependence of the waveform on the EOS can be seen more clearly by decomposing each waveform into amplitude $A(t)$ and phase $\Phi(t)$ with the relation $h_+(t) - i h_\times(t) = A(t)e^{-i\Phi(t)}$.  In Fig.~\ref{fig:hamp}, the amplitude as a function of time for each BHNS waveform is compared to a BBH waveform of the same value of $Q$ and $M_{\rm NS}$.  At early times, the waveform is almost identical to the BBH waveform.  However, a few ms before the maximum amplitude is reached, the amplitude begins to depart from the BBH case.  For each $Q$ and $M_{\rm NS}$, this departure from the BBH waveform is monotonic in $\Lambda$.  Neutron stars with large values of $\Lambda$ merge earlier, and as a result the waveforms reach a smaller maximum amplitude.  The phase of each waveform is compared to that of the EOB BBH waveform $\Phi_{\rm EOB}$ in Fig.~\ref{fig:phasetime}.  At early times the phase oscillates about the EOB phase due to initial eccentricity in the numerical waveform discussed in Sec.~\ref{sec:window}. At later times, closer to the merger, tidal interactions lead to a higher frequency orbit; this, together with correspondingly stronger gravitational wave emission, means the BHNS phase accumulates faster than the EOB phase.  This continues for 1--2~ms after the waveform reaches its maximum amplitude (indicated by the dot on each curve).  Eventually the amplitude drops significantly, and numerical errors dominate the phase.  We truncate the curves when the amplitude drops below 0.01.

\begin{figure}[!htb]
\begin{center}
\includegraphics[width=80mm]{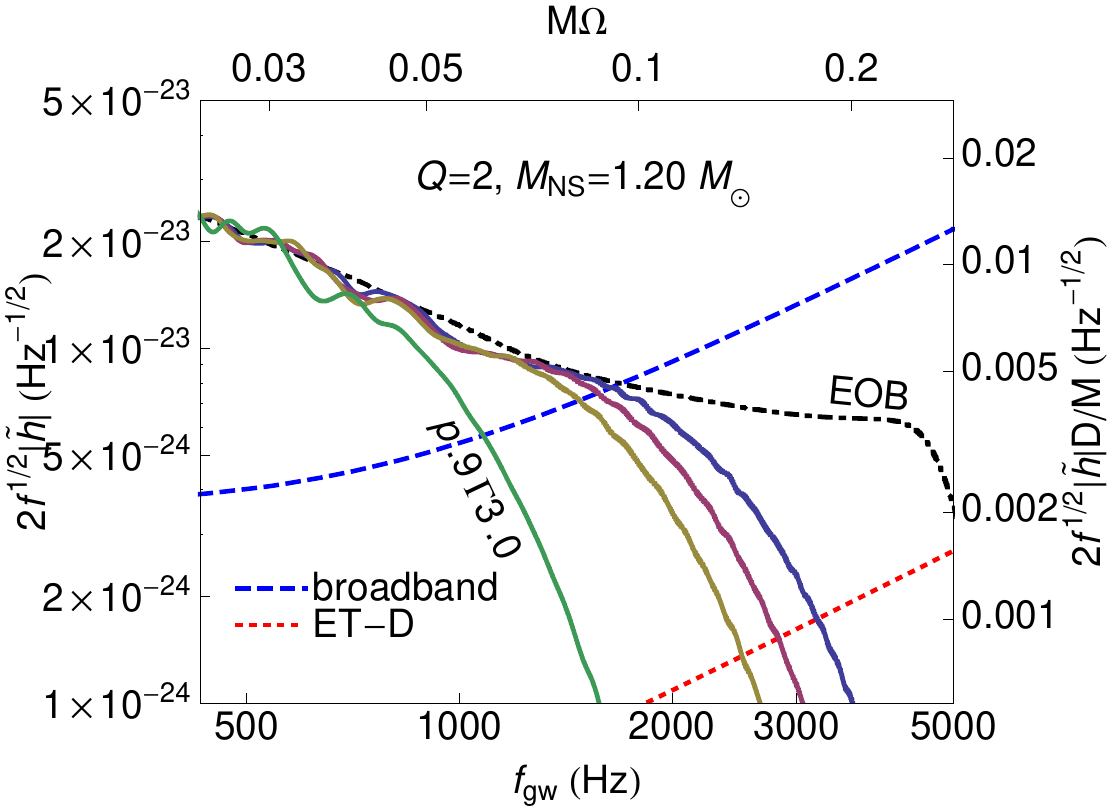}
\includegraphics[width=80mm]{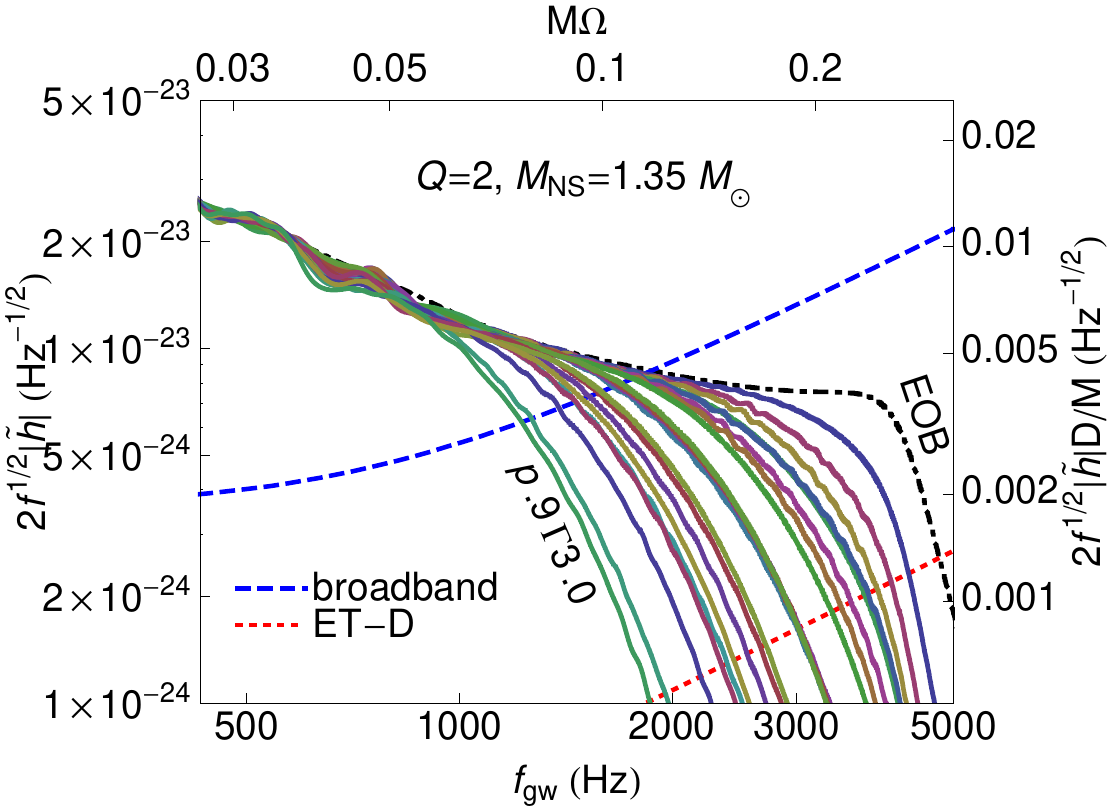}
\includegraphics[width=80mm]{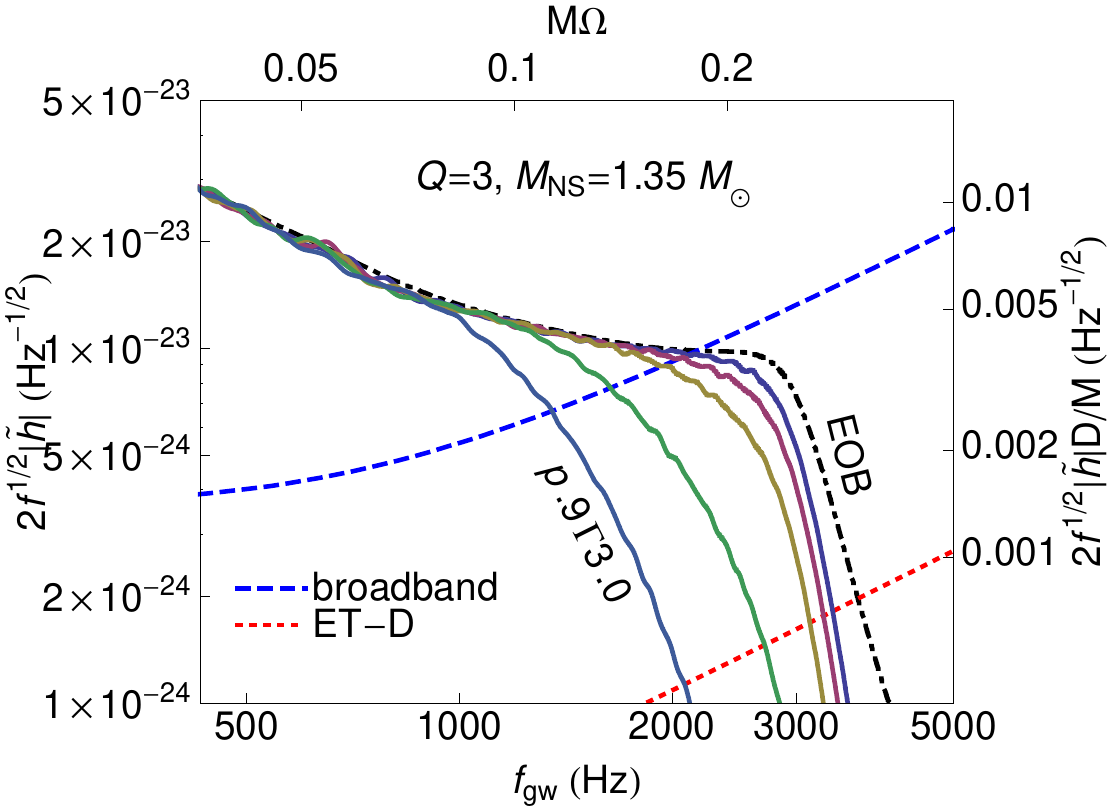}
\end{center}
\caption{ \label{fig:spectrum}
Weighted Fourier transform $2 f^{1/2} |\tilde h(f)|$ of numerical waveforms where $\tilde h = \frac{1}{2}(\tilde h_+ + \tilde h_\times)$.  Dot-dashed curves are EOB waveforms with the same $Q$ and $M_{\rm NS}$.   The left axis is scaled to a distance of 100~Mpc, and the noise $S_n^{1/2}(f)$ for broadband aLIGO and ET-D are shown for comparison.  In each plot the numerical waveform monotonically approaches the EOB waveform as the tidal parameter $\Lambda$ decreases.  Matching and splicing conventions are those of Fig.~\ref{fig:waveform}.
}
\end{figure}

\begin{figure}[!htb]
\begin{center}
\includegraphics[width=80mm]{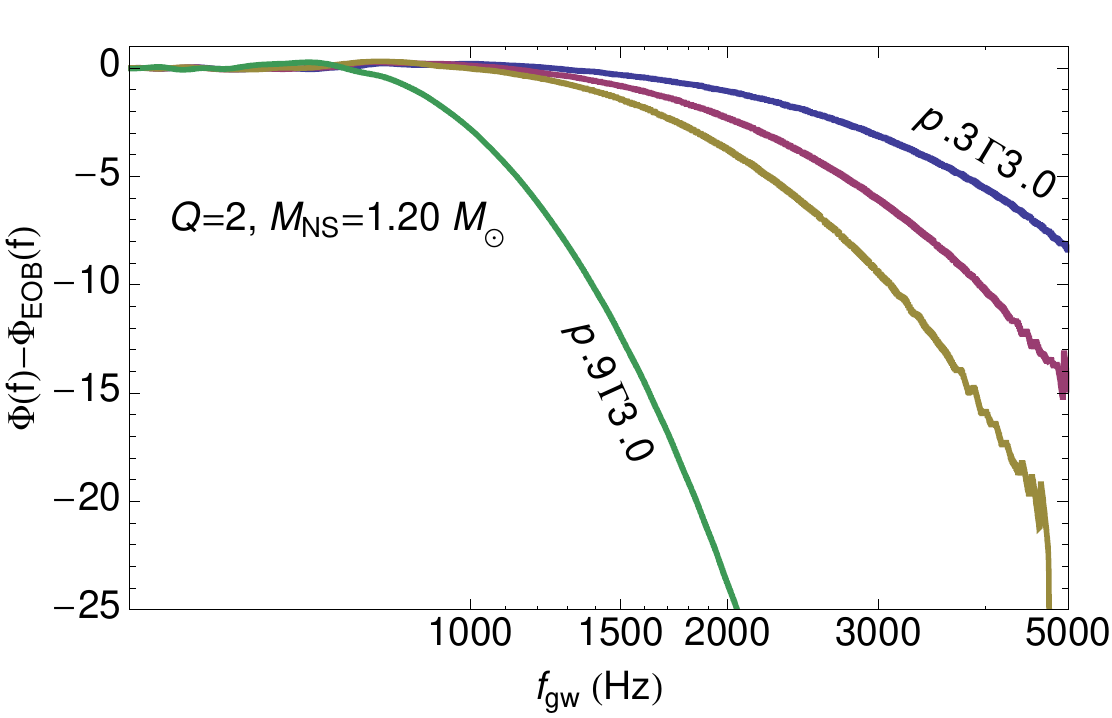}
\includegraphics[width=80mm]{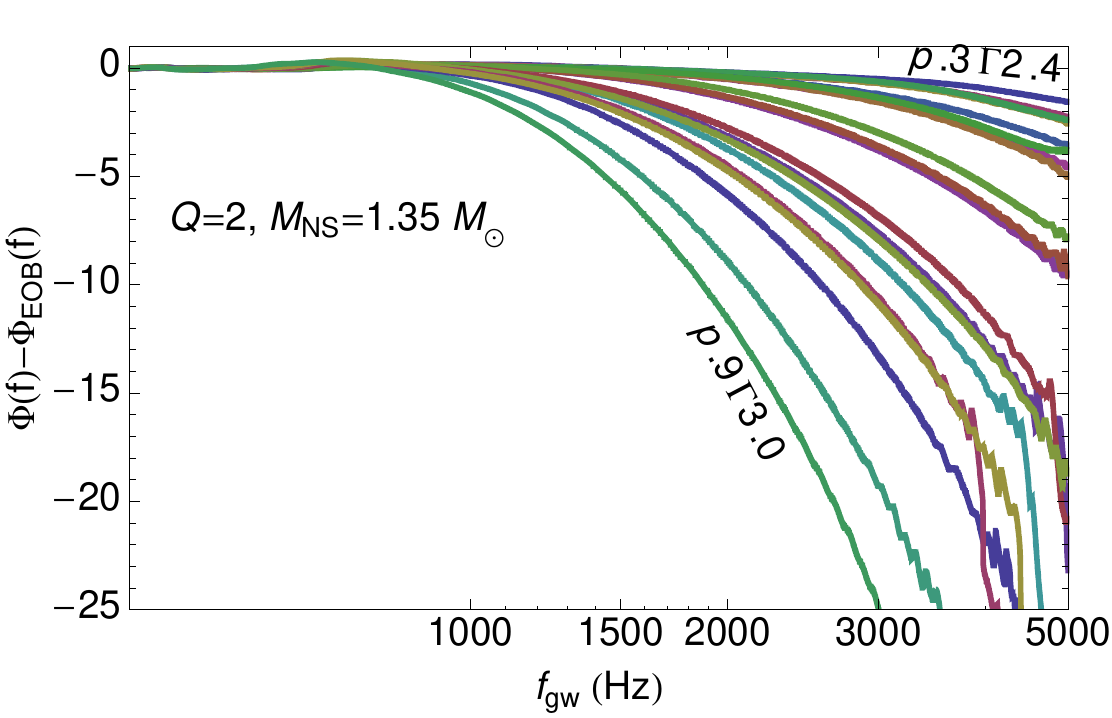}
\includegraphics[width=80mm]{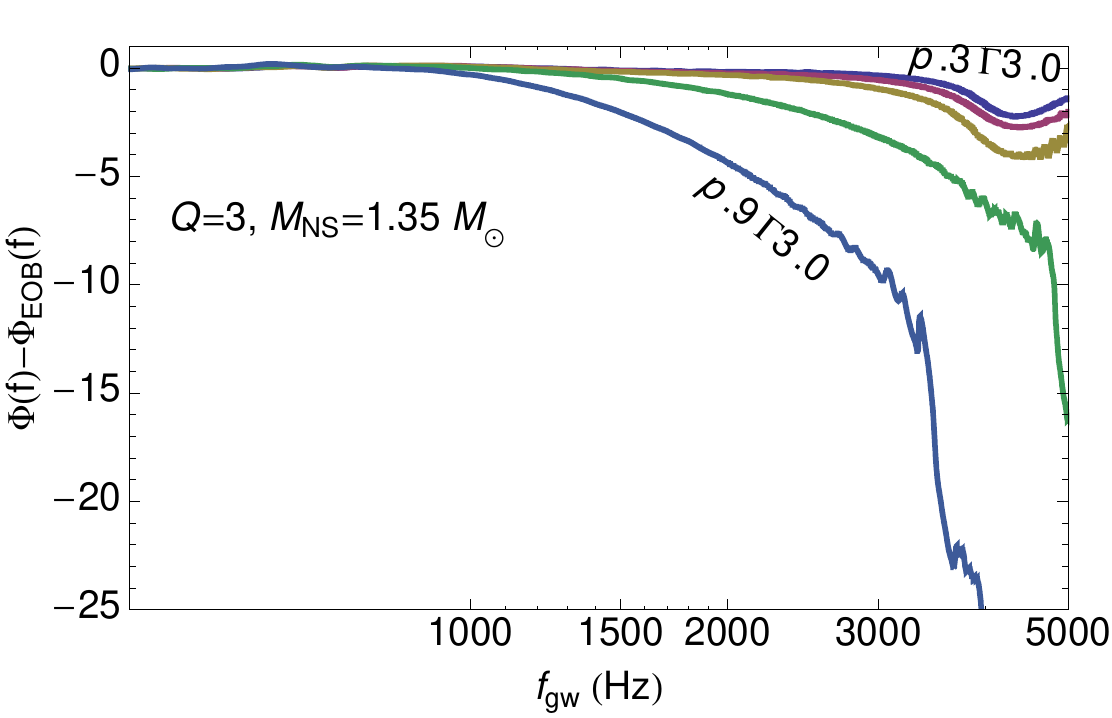}
\end{center}
\caption{ \label{fig:phasefreq}
Cumulative phase difference $\Phi-\Phi_{\rm EOB}$ of the Fourier transform between BHNS waveform and EOB BBH waveform of the same mass and mass ratio.  The phase is defined by breaking up the Fourier transform $\tilde h = \frac{1}{2}(\tilde h_+ + \tilde h_\times)$ of each waveform into amplitude and cumulative phase $\tilde h(f) = A(f)e^{-i\Phi(f)}$.  Matching and splicing conventions are those of Fig.~\ref{fig:waveform}.
}
\end{figure}

The monotonic dependence of the waveform on $\Lambda$ can again be seen in its Fourier transform $\tilde h$, shown in Figs.~\ref{fig:spectrum} and~\ref{fig:phasefreq}, which is decomposed into amplitude and phase by $\tilde h(f) = A(f)e^{-i\Phi(f)}$.  The predicted EOS dependent frequency cutoff in the waveform~\cite{Vallisneri2000} is clearly shown in the amplitude\footnote{Tidal disruption occurs after the
onset of mass shedding of the neutron star. The frequency at the onset
of mass shedding is usually much lower than that of tidal disruption for
BHNS binaries~\cite{ShibataTaniguchi2008}. In Ref.~\cite{Vallisneri2000}, mass-shedding frequency was
identified as the cutoff frequency but this underestimates the true
cutoff frequency.  See also Refs.~\cite{LaiRasioShapiro1993, LaiRasioShapiro1994} for a discussion of dynamical mass transfer.}.  Neutron stars that are more easily disrupted (larger $\Lambda$) result in an earlier and lower frequency drop in their waveform amplitude than NS with smaller $\Lambda$.  The phase $\Phi(f)$ relative to the corresponding BBH waveform has a much smoother behavior than the phase of the time domain waveform.  This feature will be useful in evaluating the Fisher matrix in Sec.~\ref{sec:errors}.  The noise that is seen at frequencies above $\sim 3000$~Hz is the result of numerical errors in the simulation and has no effect on the error estimates below.

\section{Hybrid Waveform Construction}
\label{sec:hybrid}

Since our numerical simulations typically begin $\sim$5 orbits before merger, it is necessary to join the numerical waveforms to analytic waveforms representing the earlier inspiral. There is a substantial literature comparing analytic and phenomenological waveforms with numerical waveforms extracted from simulations of BBH coalescence. For example, it has been shown that the 3.5 post-Newtonian (TaylorT4) waveform agrees well with equal mass BBH waveforms up to the last orbit before merger~\cite{BoyleBrownKidder2007}. For unequal mass systems, the EOB formalism (see Ref.~\cite{DamourNagar2009review} for a review) has proven to be a powerful tool to generate analytic waveforms that agree with numerical simulations. Free parameters in the EOB formalism have been fit to numerical BBH waveforms to provide analytic (phenomenological) waveforms that extend to the late, non-adiabatic inspiral as well as the ringdown. These EOB waveforms appear to be in good agreement with numerical BBH waveforms for mass ratios at least up to $Q=4$~\cite{DamourNagar2009}.  Although we have not explored them in this context, other approaches have also been taken for constructing phenomenological inspiral-merger-ringdown waveforms~\cite{Ajith2008, Ajith2011, Santamaria2010, Sturani2010a, Sturani2010b}.

For equal-mass BNS, Read et al.~\cite{ReadMarkakisShibata2009} compared the numerical BNS waveform during inspiral to a point particle post-Newtonian waveform. Specifically, they used the 3.5 post-Newtonian (TaylorT4) waveforms matched on to the numerical waveforms to study the measurability of EOS parameters. They found that differences between the analytic and numerical waveforms become apparent $4-8$ cycles before the post-Newtonian coalescence time.  

The leading and post-1-Newtonian quadrupole tidal effects have recently been incorporated into the post-Newtonian formalism and used to compute corrections to the point-particle gravitational waveforms~\cite{FlanaganHinderer2008, VinesFlanagan2010, VinesFlanaganHinderer2011}.  These post-Newtonian contributions along with a fit to the 2PN tidal contribution have also been incorporated into the EOB formalism and compared to long simulations ($\sim20$ GW cycles), where they find agreement with the simulations to $\pm0.15$~rad over the full simulation up to merger~\cite{BaiottiDamour2011}.

For the BHNS systems discussed here, we have matched the numerical waveforms to EOB waveforms that include inspiral, merger, and ringdown phases instead of post-Newtonian waveforms which are often not reliable during the last few cycles for higher mass ratios.
This choice also allows us to use longer matching windows that average over numerical noise and the effects of eccentricity as shown in Sec.~\ref{sec:window}. We have chosen to use the EOB formalism to generate inspiral-merger-ringdown waveforms, although we note that other phenomenological waveforms would probably work.  For simplicity, and because it appears that an accurate description of the late inspiral dynamics just before merger requires 2PN tidal corrections~\cite{DamourNagar2010, BaiottiDamour2010, BaiottiDamour2011} which are not yet known, we will use the EOB waveforms without tidal corrections. Our results will therefore be lower limits on the measurability of EOS parameters since the EOS dependence is coming solely from the numerical waveforms. 

\subsection{Matching procedure}
\label{sec:match}

We use a method similar to that described by Read et al.~\cite{ReadMarkakisShibata2009} to join each of the numerical BHNS waveforms to a reference EOB waveform, generating a hybrid EOB--numerical waveform.  Denote a complex numerical waveform by $h_{\rm NR}(t) = h_+^{\rm NR}(t)-ih_\times^{\rm NR}(t)$ and an EOB waveform with the same $Q$ and $M_{\textrm{NS}}$ by $h_{\rm EOB}(t) = h_+^{\rm EOB}(t)-ih_\times^{\rm EOB}(t)$.  A constant time-shift $\tau$ and phase-shift $\Phi$ can be applied to the EOB waveform to match it to a section of the numerical waveform by rewriting it as $h_{\rm EOB}(t-\tau)e^{-i\Phi}$.  We hold the numerical waveform fixed because we must specify a matching window $T_I<t<T_F$, and as discussed below, there is only a small region of the numerical waveforms over which a valid match can be performed.  Once the values of $\tau$ and $\Phi$ are determined, we will then choose to instead hold the EOB waveform fixed and shift the numerical waveform in the opposite direction by rewriting it as $h_{\rm NR}^{\rm shift}(t) = h_{\rm NR}(t+\tau)e^{+i\Phi}$.  This is done so that all of the numerical waveforms with the same $Q$ and $M_{\textrm{NS}}$ are aligned relative to a single fixed reference EOB waveform.

Over a matching window $T_I<t<T_F$ (bounded by solid vertical lines in Fig.~\ref{fig:waveform}), the normalized match between the waveforms is defined as 
\begin{equation}
\label{eq:match}
m(\tau, \Phi) = \frac{ \textrm{Re}\, \left[ z(\tau) e^{i\Phi} \right] 
}{
\sigma_{\rm NR}\sigma_{\rm EOB}(\tau)
},
\end{equation}
where
\begin{equation}
z(\tau) = 
\int_{T_I}^{T_F}h_{\rm NR}(t)h_{\rm EOB}^*(t-\tau) \,dt
\end{equation}
and the normalizations for each waveform in the denomenator are defined as 
\begin{equation}
\sigma_{\rm NR}^2 = \int_{T_I}^{T_F} |h_{\rm NR}(t)|^2\,dt
\end{equation}
and
\begin{equation}
\sigma_{\rm EOB}^2(\tau) = \int_{T_I}^{T_F} |h_{\rm EOB}(t-\tau)|^2\,dt.
\end{equation}
The time-shift $\tau$ and phase $\Phi$ are chosen to maximize the match $m(\tau, \Phi)$ for a fixed matching window. Explicitly, the phase is determined analytically to be $\Phi = -\arg[z(\tau)]$; plugging this result back into Eq.~(\ref{eq:match}), the time-shift is given by maximizing $|z(\tau)|/[ \sigma_{\rm NR}\sigma_{\rm EOB}(\tau) ]$ over $\tau$.  As stated above, once $\tau$ and $\Phi$ are found we shift the numerical waveform in the opposite direction to generate $h_{\rm NR}^{\rm shift}(t) = h_{\rm NR}(t+\tau)e^{+i\Phi}$.

A hybrid waveform is generated by smoothly turning off the EOB waveform and smoothly turning on the shifted numerical waveform over a splicing window $S_I<t<S_F$ (bounded by dotted vertical lines in Fig.~\ref{fig:waveform}) which can be chosen independently of the matching window.  As in Ref.~\cite{ReadMarkakisShibata2009}, we employ Hann windows
\begin{eqnarray}
w_{\rm off}(t) &=& \frac{1}{2}\left[1+\cos\left(\frac{\pi [t - S_I]}{S_F - S_I}\right)\right]\\
w_{\rm on}(t) &=& \frac{1}{2}\left[1-\cos\left(\frac{\pi [t - S_I]}{S_F - S_I}\right)\right].
\end{eqnarray}
The hybrid waveform is then written
\begin{equation}
h_{\rm hybrid}(t) =\
\left\{\begin{array}{lc}
h_{\rm EOB}(t) & \, t<S_I \\
w_{\rm off}(t) h_{\rm EOB}(t) + w_{\rm on}(t) h_{\rm NR}^{\rm shift}(t) & \, S_I<t<S_F \\
h_{\rm NR}^{\rm shift}(t) & \, t>S_F
\end{array}\right..
\end{equation}

As shown in Fig.~\ref{fig:waveform}, we choose the start of the splicing interval to be the same as the start of the matching window $S_I = T_I$ and choose the end of the splicing window to be $S_F = T_I+4$~ms.  It is also necessary to use these windows to smoothly turn on the hybrid waveform at low frequency when performing a discrete Fourier transform to avoid the Gibbs phenomenon.  Unlike the case for BNS waveforms, it is not necessary to window the end of the hybrid waveform as the amplitude rapidly decays to zero anyway during the ringdown.

For concreteness we define $t=0$ as the EOB BBH merger time $t_M^{\rm EOB}$ when the EOB waveform reaches its maximum amplitude.  After matching to the EOB waveform, the time when the numerical BHNS waveform reaches its maximum amplitude is $t_M^{\rm NR}$.

\subsection{Dependence on matching window}
\label{sec:window}

Because the numerical BHNS waveforms are close but not identical to the EOB BBH waveform during the inspiral and because there is some noise in the BHNS waveforms, the time shift that maximizes the match depends on the choice of matching window.  The matching window should exclude the first couple of cycles of the numerical waveform during which time the simulation is settling down from the initial conditions.  It should also exclude the merger/ringdown which are strongly dependent on the presence of matter. The window must also be wide enough to average over numerical noise and, as we shall see below, the effects of eccentricity in the simulations.  

\begin{figure}[!htb]
\begin{center}
\includegraphics[width=80mm]{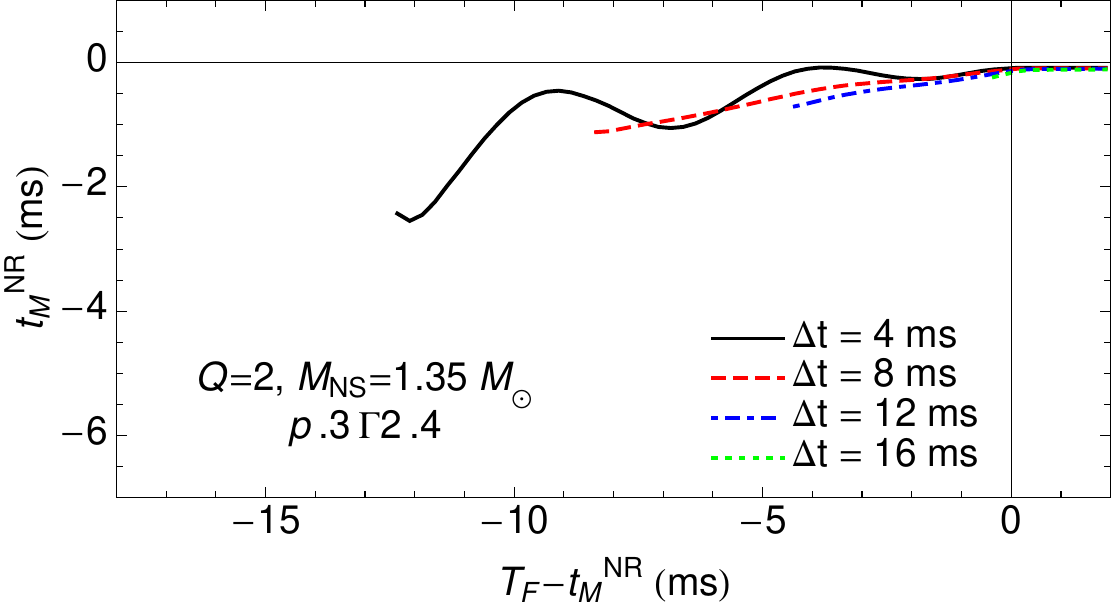}
\includegraphics[width=80mm]{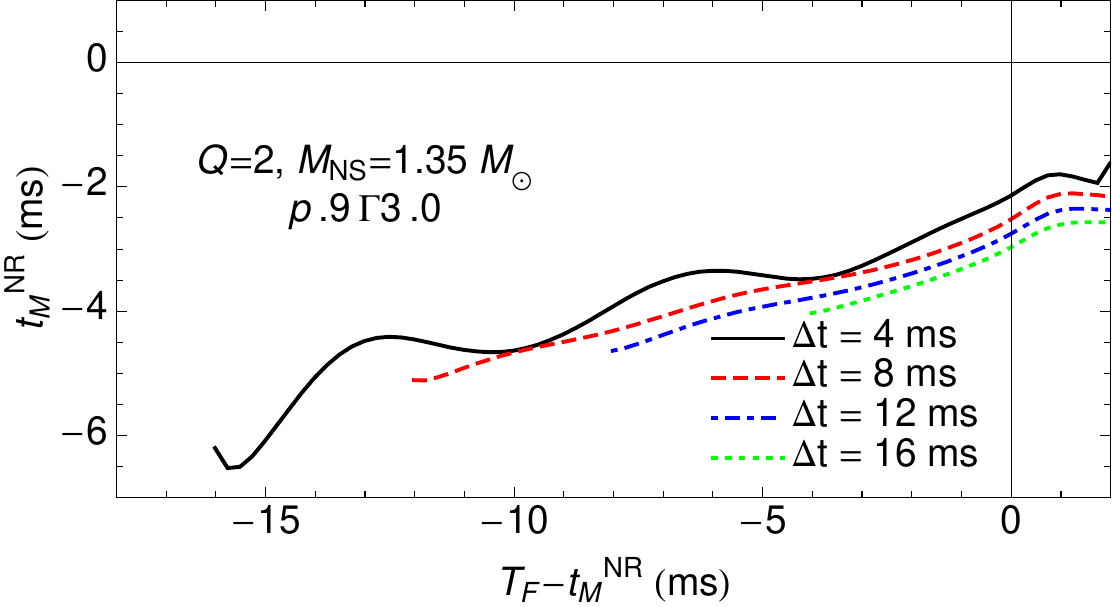}
\end{center}
\caption{ \label{fig:timeshift}
Dependence of time shift between numerical and EOB waveform on the end time $T_F-t_M^{\rm NR}$ and width $\Delta t$ of the matching window.  $Q=2$ and $M_{\rm NS}=1.35M_\odot$ for each waveform.  The EOS used are p.3$\Gamma$2.4 (top panel),  and p.9$\Gamma$3.0 (bottom panel).  The EOB waveform has zero eccentricity.
}
\end{figure}

The numerical merger time $t_M^{\rm NR}$ relative to the EOB BBH merger time $t_M^{\rm EOB}$ as a function of the end of the matching window $T_F-t_M^{\rm NR}$ provides a useful diagnostic of the matching procedure.
Results for matching two $Q=2, M_{\rm NS} = 1.35 M_\odot$ waveforms with different equations of state to an EOB waveform are shown in Fig.~\ref{fig:timeshift}.  The horizontal axis is the end time $T_F$ of the matching window relative to the numerical merger time $t_M^{\rm NR}$.  For negative values, the matching window contains the BHNS inspiral only.  For positive values, the matching window also contains part of the BHNS ringdown.  The vertical axis is the location of the shifted numerical merger time $t_M^{\rm NR}$ after finding the best match.  Four different window durations $\Delta t = T_F - T_I$ are shown.  The drift in the best fit merger time $t_M^{\rm NR}$ most likely arises from the neglect of tidal effects in the EOB waveform which lead to an accumulating phase shift in the waveform, although it could also arise from numerical angular momentum loss from finite resolution of the simulations. Further work is in progress to understand this issue~\cite{Read2011}.

When the matching window duration is of order one orbital period or shorter, the time-shift oscillates as a function of $T_F-t_M^{\rm NR}$.  We attribute this effect to the eccentricity in the numerical waveform that results from initial data with no radial velocity.  For larger matching-window durations, the effect of eccentricity is averaged out.

\begin{figure}[!htb]
\begin{center}
\includegraphics[width=80mm]{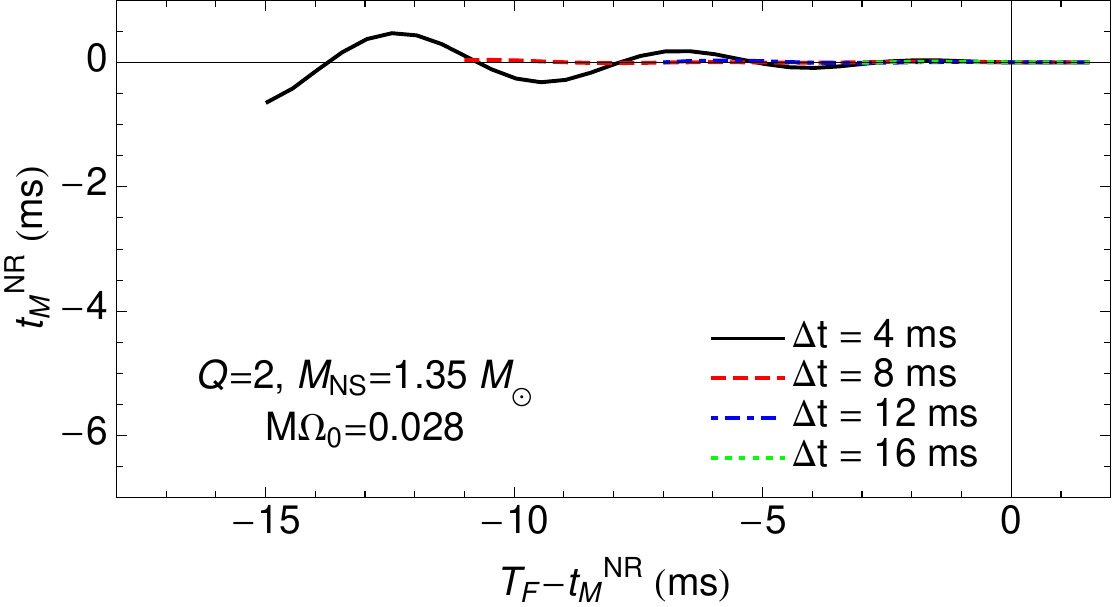}

\end{center}
\caption{ \label{fig:eccentric}
Same as Fig.~\ref{fig:timeshift}, but matching an eccentric EOB BBH waveform with the quasicircular initial condition $M\Omega_0=0.028$ to a zero eccentricity EOB BBH waveform.
}
\end{figure}

To demonstrate concretely that the decaying oscillations for $\Delta t=4$~ms are the result of eccentricity, we matched an EOB BBH waveform with eccentricity to the equivalent zero eccentricity EOB BBH waveform.  EOB waveforms can be generated with small eccentricity by starting the EOB equations of motion with quasicircular (zero radial velocity) initial conditions late in the inspiral.  The result is shown in Fig.~\ref{fig:eccentric} for an EOB waveform with the same quasicircular initial conditions as the simulation for the EOS p.3$\Gamma$2.4 shown in Fig.~\ref{fig:timeshift}.  The oscillations take exactly the form of those shown in Fig.~\ref{fig:timeshift}, except without the drift and offset.  

We estimate that the initial eccentricities in the simulations used in this paper are $e_0\sim0.03$.  Decreasing the initial eccentricity by about an order of magnitude, possibly using an iterative method that adjusts the initial radial velocity~\cite{Pfeiffer2007}, will remove this issue and allow one to determine the phase shift due to tidal interactions during the inspiral part of the simulation.

\section{Parameter estimation}
\label{sec:errors}

The output of a gravitational-wave detector $s(t) = n(t) + h(t)$ is the sum of detector noise $n(t)$ and a possible gravitational-wave signal $h(t)$. Stationary, Gaussian noise is characterized by its power spectral density (PSD) $S_n(|f|)$ defined by
\begin{equation}
\langle \tilde{n}(f) \tilde{n}^\ast(f') \rangle = \frac{1}{2} \delta(f-f') S_n(|f|) \; . 
\end{equation}
The gravitational wave signal is given in terms of the two polarizations of the gravitational wave by
\begin{equation}
h(t) = F_+ h_+(t) + F_\times h_\times(t),
\end{equation}
where $F_{+,\times}$ are the detector response functions and depend on the location of the binary and the polarization angle of the waves.  We assume the binary is optimally located at the zenith of the detector and optimally oriented with its orbital plane parallel to that of the detector.  This condition is equivalent to averaging $h_+$ and $h_\times$ ($F_+ = F_\times = 1/2$).

It is well known~\cite{wainstein:1962} that the optimal statistic for detection of a known signal $h(t)$ in additive Gaussian noise is 
\begin{equation}
\mathcal{\rho} = \frac{(h|s)}{\sqrt{(h|h)}}
\end{equation}
where the inner product between two signals $h_1$ and $h_2$ is given by
\begin{equation}
(h_1|h_2)=4{\rm Re} \int_0^\infty \frac{\tilde h_1(f) \tilde h_2^*(f)}{S_n(f)}\,df.
\end{equation}
In searches for gravitational-wave signals from compact binary mergers, a parametrized signal $h(t;\theta^A)$ is known in advance of detection, and the parameters $\theta^A$ must be estimated from the measured detector output $s(t)$. The parameters $\theta^A$ of an inspiral are estimated by maximizing the inner product of the signal $s(t)$ over the template waveforms $h(t;\theta^A)$. In the high signal-to-noise limit, the statistical uncertainty in the estimated parameters $\hat{\theta}^A$ arising from the instrumental noise can be estimated using the Fisher matrix
\begin{equation}
\label{eq:fisher}
\Gamma_{AB} = \left. \left(\frac{\partial h}{\partial \theta^A}\left|\frac{\partial h}{\partial \theta^B}\right.\right) \right|_{\hat{\theta}^A} \; .
\end{equation}
Note that $\hat{\theta}^A$ are the parameter values that maximize the signal-to-noise. 
The variance $\sigma_A^2 = \sigma_{AA}= \langle(\Delta\theta^A)^2\rangle$ and covariance $\sigma_{AB} = \langle\Delta\theta^A\Delta\theta^B\rangle$ of the parameters are then given in terms of the Fisher matrix by
\begin{equation}
\langle\Delta\theta^A\Delta\theta^B\rangle = (\Gamma^{-1})^{AB}.
\end{equation}

For hybrid waveforms, the partial derivatives in the Fisher matrix must be approximated with finite differences. It is most robust to compute the derivatives of the Fourier transforms used in the inner product. We rewrite the Fourier transform of each waveform in terms of the amplitude $A$ and phase $\Phi$ as $\exp[ \ln A - i \Phi]$ as given in Eq.~(\ref{eq:logampphase}). The derivatives $\partial \ln A / \partial \theta^A$ and $\partial \Phi / \partial \theta^A$ are then evaluated with finite differencing.  More details of this and the other methods we tested are given in Appendix~\ref{app:numeval}.

In general, errors in the parameters $\theta^A$ are correlated with each other forming an error ellipsoid in parameter space determined by the Fisher matrix $\Gamma_{AB}$. The uncorrelated parameters that are best extracted from the signal are found by diagonalizing $\Gamma_{AB}$. These new parameters are linear combinations of the original parameters $\theta^A$.  We focus attention below on the two parameters $\log(p_1)$ and $\Gamma$, and fix all other parameters as follows.  We use the masses and spins determined from the numerical simulations and fix the time and phase shifts as determined during the hybrid waveform construction.  We therefore construct the error ellipses in $\{\log(p_1), \Gamma\}$ parameter space and identify the parameter with the smallest statistical errors.  We will leave an analysis of correlations due to uncertainty in masses and BH spin to future work.

\subsection{Broadband aLIGO and ET}

\begin{figure}[!htb]
\begin{center}
\includegraphics[width=80mm]{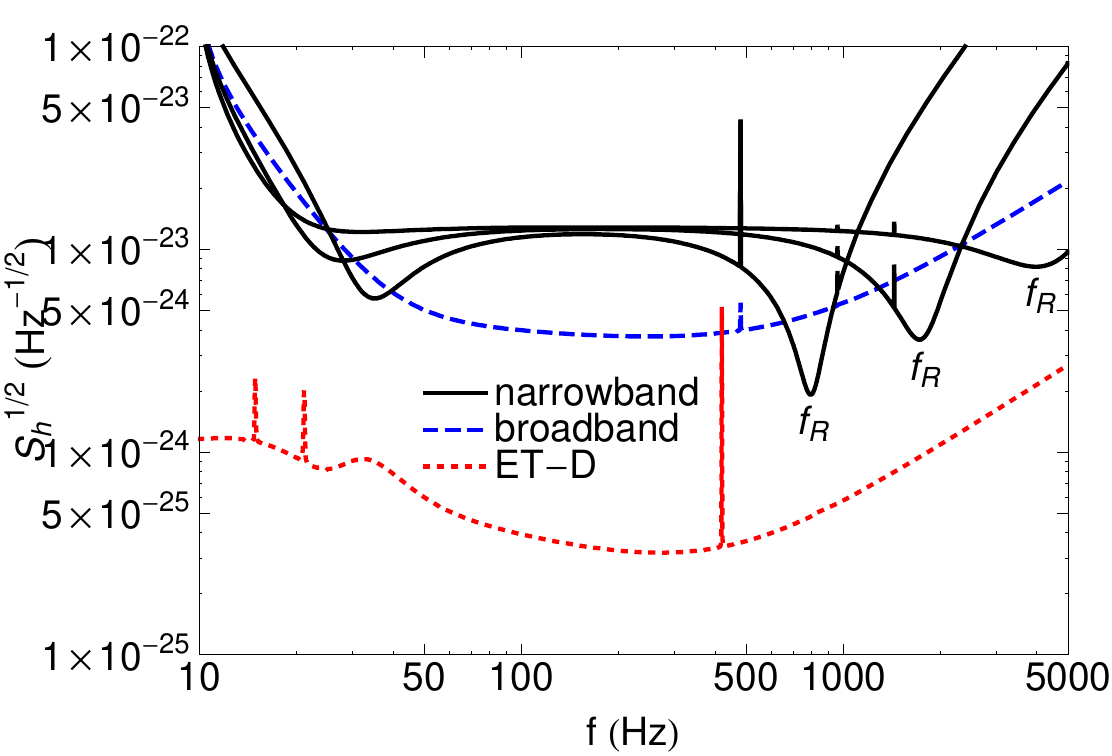}
\end{center}
\caption{ \label{fig:noisecurves}
Noise PSD for broadband aLIGO (dashed blue), ET-D (dotted red) and various configurations of narrowband aLIGO (solid black).  The minima of the narrowband configuration are labeled $f_R$
}
\end{figure}

For the BHNS systems discussed here, the greatest departure from BBH behavior occurs for gravitational-wave frequencies in the range 500--5000~Hz. As a result, detector configurations optimized for detection of BHNS systems with low noise in the region below 500~Hz are not ideal for estimating EOS parameters.  We therefore present results for the broadband aLIGO noise curve~\cite{LIGOnoise} and the ET-D noise curve~\cite{ETDnoise} shown in Fig.~\ref{fig:noisecurves}.  The broadband aLIGO configuration uses zero-detuning of the signal recycling mirror and a high laser power, resulting in significantly lower noise above 500~Hz at the expense of slightly higher noise at lower frequencies.  Several noise curves have been considered for the Einstein Telescope denoted ET-B~\cite{ETBnoise}, ET-C~\cite{ETCnoise}, and ET-D~\cite{ETDnoise}.  We will use the most recent ET-D configuration, and note that in the 500--5000~Hz range all of the ET configurations have a similar sensitivity.  The published noise curves, and those used in this paper, are for a single interferometer of 10~km with a 90$^\circ$ opening angle.  The current ET proposal is to have three individual interferometers each with a 60$^\circ$ opening angle.  This will shift the noise curve down appoximately 20\%~\cite{ETDnoise}.

\begin{figure}[!htb]
\begin{center}
\includegraphics[width=80mm]{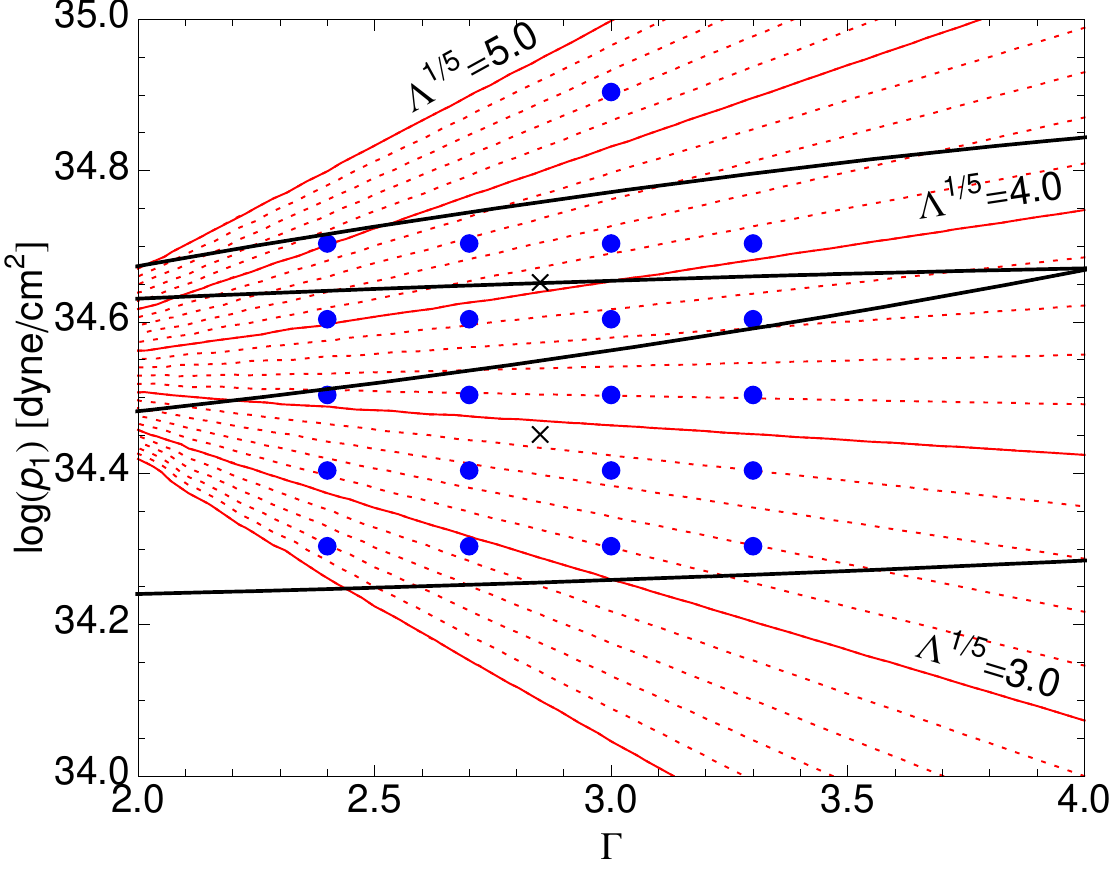}
\end{center}
\caption{ \label{fig:ellipsebroad}
Two 1--$\sigma$ error ellipses for broadband aLIGO.  Evenly spaced contours of constant $\Lambda^{1/5}$ are also shown.  Each ellipse is centered on the estimated parameter $\hat\theta^A$ denoted by a $\times$.  The semimajor axes are significantly longer than the width of the figure, so each ellipse appears as a pair of parallel lines.  Matching and splicing conventions are those of Fig.~\ref{fig:waveform}.
}
\end{figure}

\begin{figure}[!htb]
\begin{center}
\includegraphics[width=80mm]{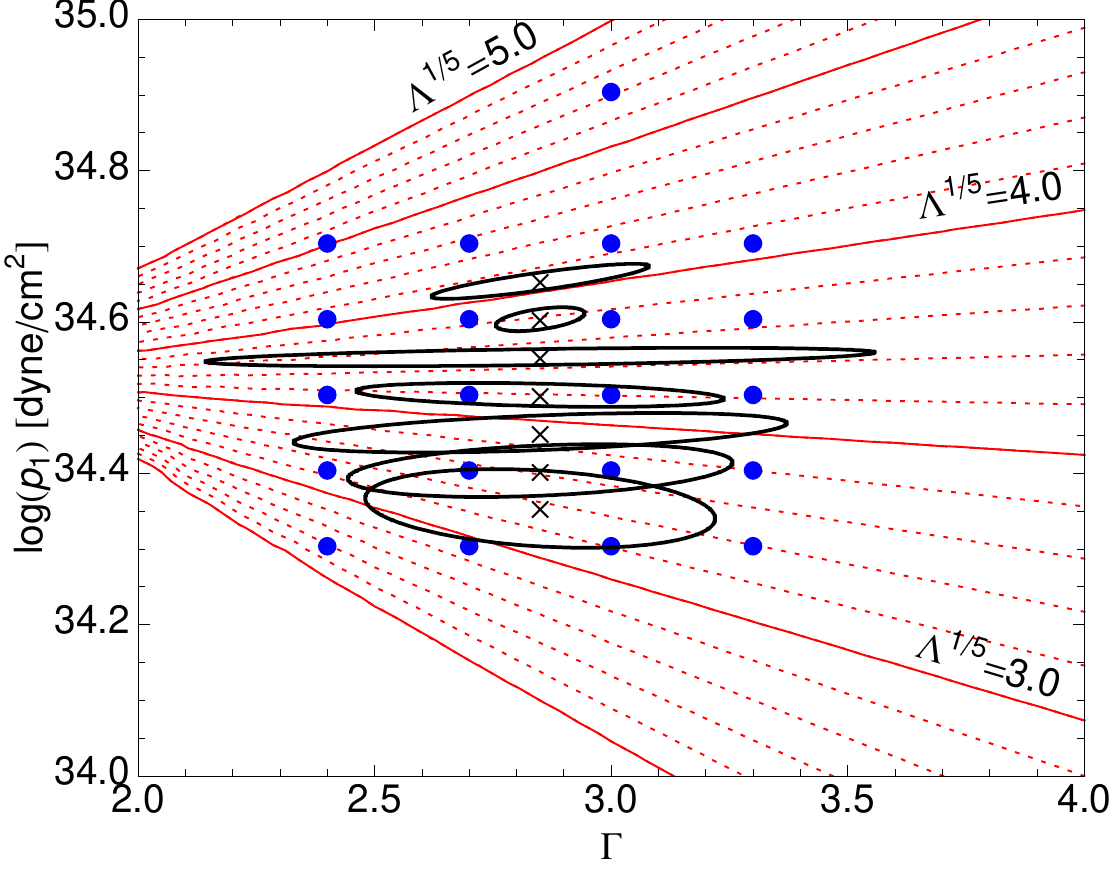}
\includegraphics[width=80mm]{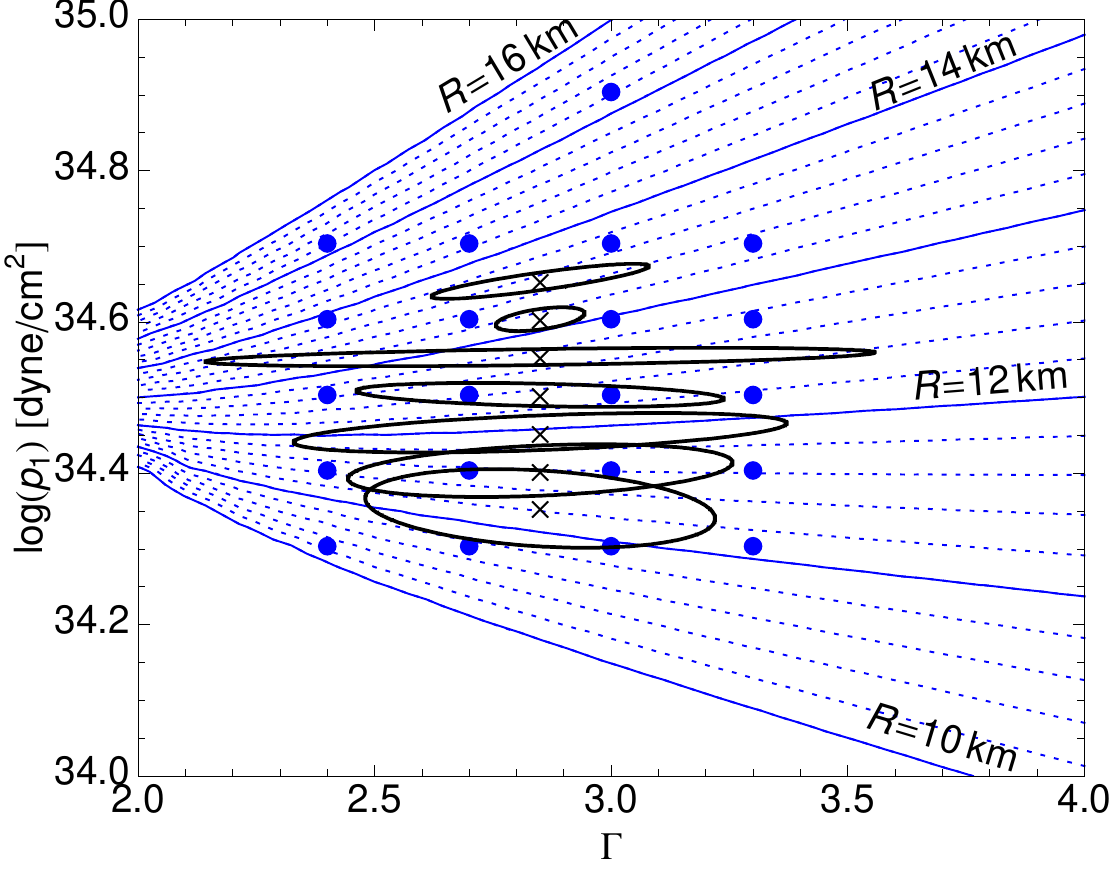}
\end{center}
\caption{ \label{fig:ellipse}
1--$\sigma$ error ellipses for ET-D.  Evenly spaced contours of constant $\Lambda^{1/5}$ ($R$) are also shown on top (bottom).  Matching and splicing conventions are those of Fig.~\ref{fig:waveform}.
}
\end{figure}

In Figs.~\ref{fig:ellipsebroad} and~\ref{fig:ellipse}, we show the resulting 1-$\sigma$ error ellipses in the 2-dimensional parameter space $\{\log(p_1), \Gamma\}$ for an optimally oriented BHNS with $Q=2$ and $M_{\rm NS}=1.35 M_\odot$ at a distance of 100~Mpc. 
Surfaces of constant $\Lambda^{1/5}$ and NS radius, which are almost parallel to each other, are also shown.  One can see that the error ellipses are aligned with these surfaces. This indicates that, as expected, $\Lambda^{1/5}$ is the parameter that is best extracted from BHNS gravitational-wave observations. Because $\Lambda^{1/5}$ and $R$ are so closely aligned we will use these two parameters interchangeably. 

As mentioned above, there is some freedom in construction of the hybrid waveforms. The size and orientation of the error ellipses also depend on the details of this construction.  We find that as long as the matching window is longer than approximately four gravitational-wave cycles to average out the effects of eccentricity and does not include the first two gravitational-wave cycles, the orientation of the error ellipses does not change significantly.  As expected, the size of the ellipses decreases as more of the numerical waveform is incorporated into the hybrid waveform. We therefore adopt the last 12~ms before merger of each numerical waveform as the matching window and the first 4~ms of the matching window for splicing as shown in Fig.~\ref{fig:waveform}.

We have emphasized that, to within present numerical
accuracy, the late-inspiral waveform is determined by the single 
parameter $\Lambda^{1/5}$.  This implies that, by using  
countours of constant $\Lambda$ in the EOS space, one could have 
obtained the constraint on the EOS, summarized in Figs.~\ref{fig:ellipsebroad} and~\ref{fig:ellipse}
by varying only a single EOS parameter.  For the simulations with other mass ratios and neutron star masses, we have used as our single parameter $\log(p_1)$ and not $\Gamma$ because countours of constant $p_1$ more closely coincide with contours of constant $\Lambda$ and because $\Lambda$ is a one to one function of $\log(p_1)$ throughout the parameter space.  The one-parameter Fisher matrix can then be evaluated with finite differencing using the waveforms and values of $\Lambda$ at two points in EOS parameter space with different $\log(p_1)$.

\begin{figure}[!htb]
\begin{center}
\includegraphics[width=80mm]{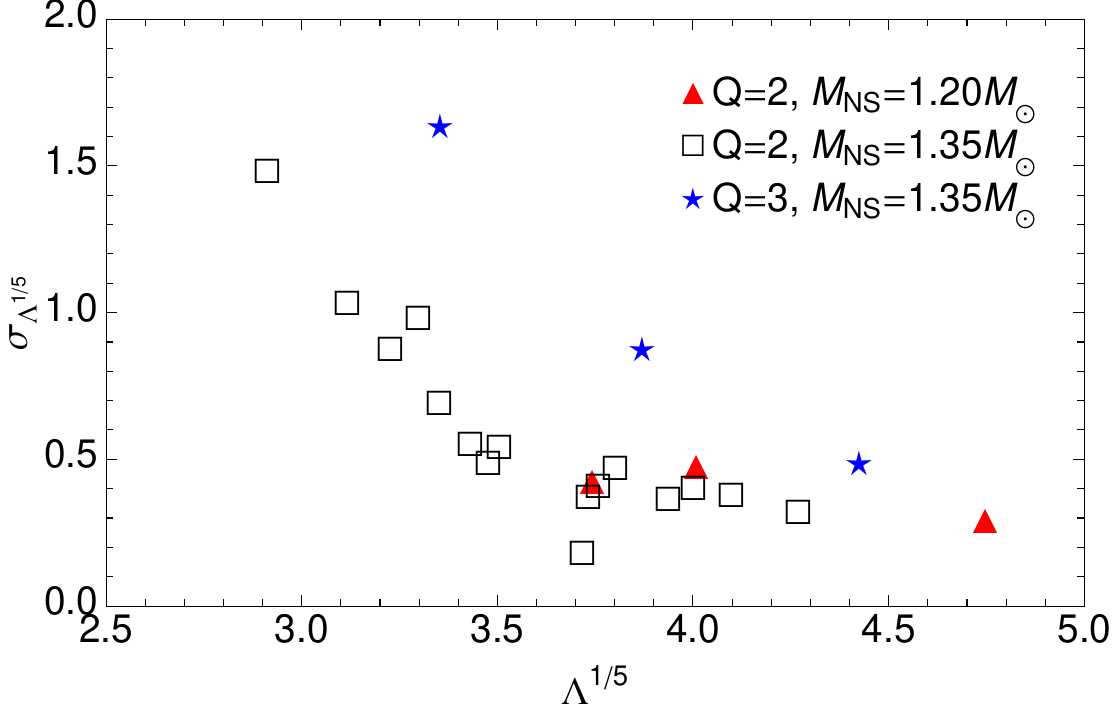}
\includegraphics[width=80mm]{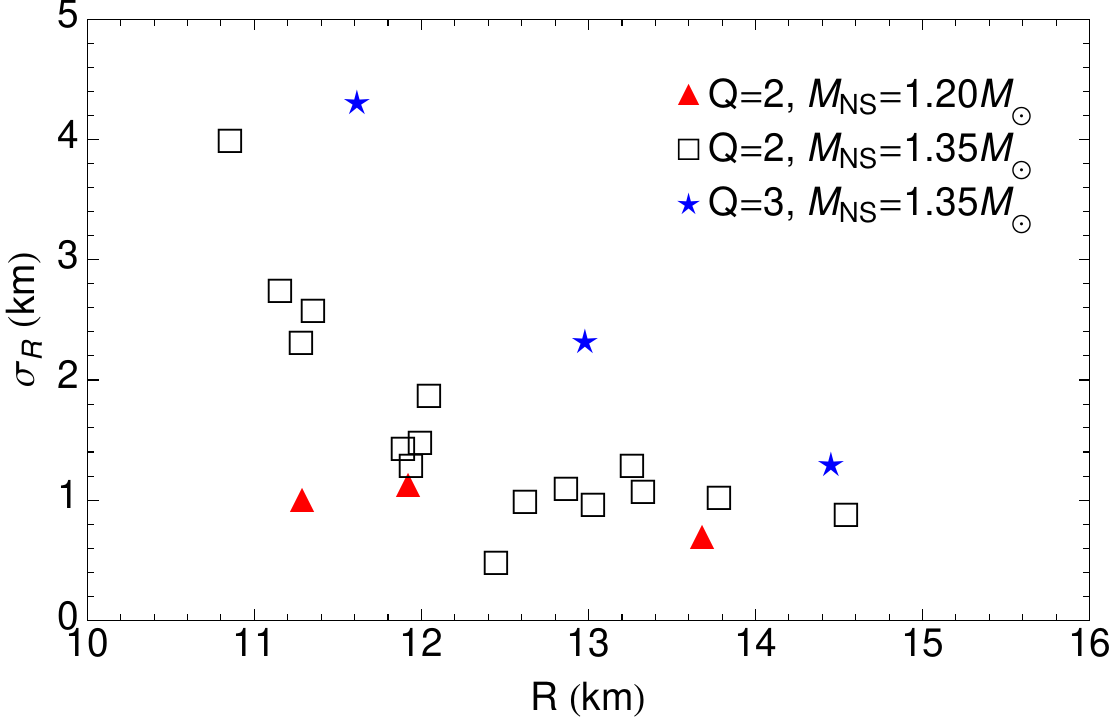}
\end{center}
\caption{ \label{fig:oneparambroad}
1-$\sigma$ uncertainty $\sigma_{\Lambda^{1/5}}$ and $\sigma_R$ as a function of the parameters $\Lambda^{1/5}$ or $R$ for the broadband aLIGO noise PSD.  Matching and splicing conventions are those of Fig.~\ref{fig:waveform}.
}
\end{figure}

\begin{figure}[!htb]
\begin{center}
\includegraphics[width=80mm]{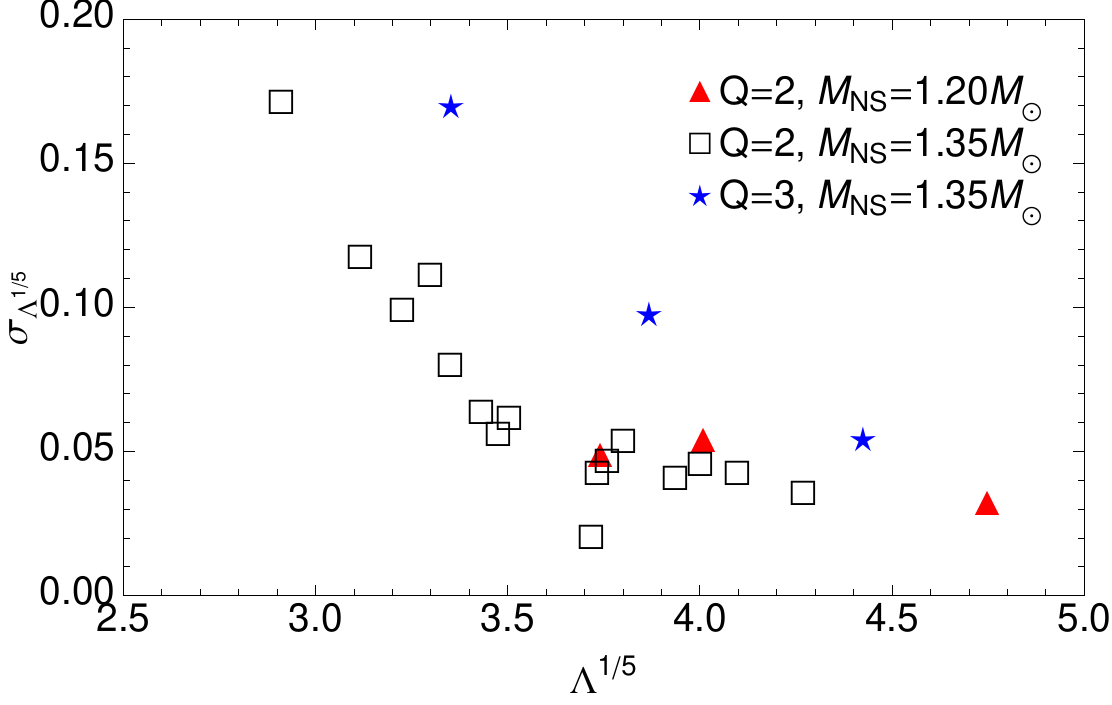}
\includegraphics[width=80mm]{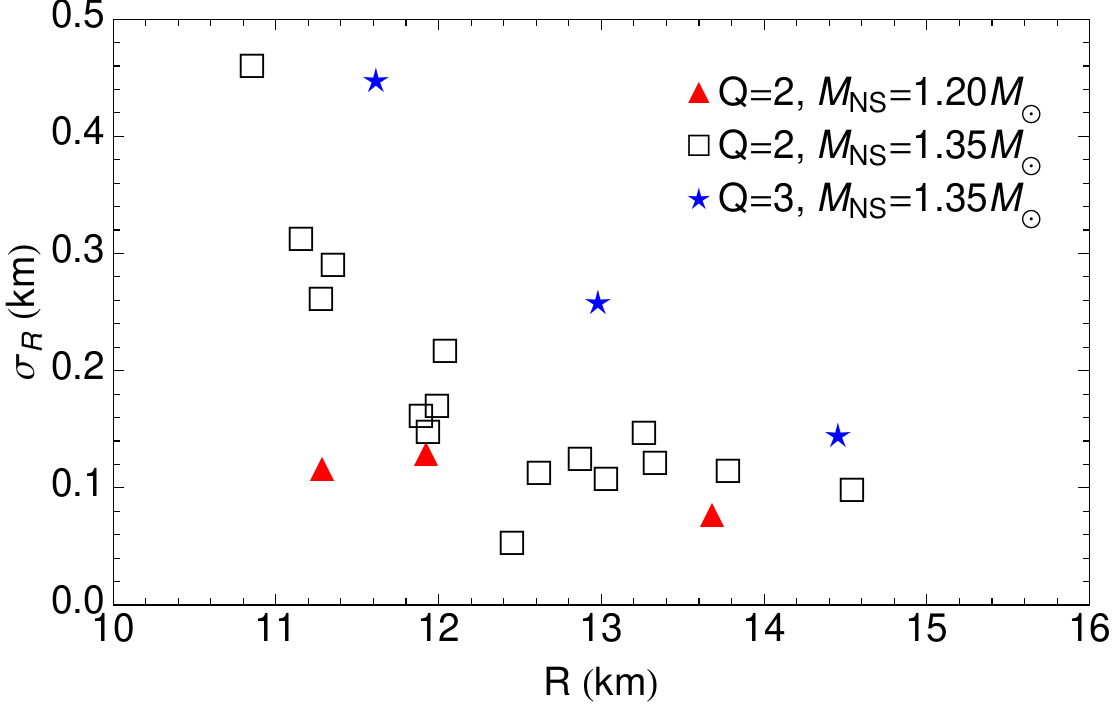}
\end{center}
\caption{ \label{fig:oneparamET}
Same as Fig.~\ref{fig:oneparambroad}, but with the ET-D noise PSD.  Error estimates are an order of magnitude smaller than for broadband aLIGO.
}
\end{figure}

The uncertainties in $\Lambda^{1/5}$ and $R$ are shown in Figs.~\ref{fig:oneparambroad} and \ref{fig:oneparamET} for broadband aLIGO and for ET respectively.  The uncertainty in these quantities is $\sim10$--40\% for broadband aLIGO and $\sim1$--4\% for ET-D. The uncertainties for the higher mass ratio $Q=3$ are somewhat larger than for $Q=2$, but not significantly so.  It is not clear how rapidly the uncertainty in $\Lambda^{1/5}$ and $R$ will increase as the mass ratio is increased toward more realistic values.  On the one hand the tidal distortion is likely to be much smaller for larger $Q$.  On the other hand the overall signal will be louder, and the merger and ringdown will occur at lower frequencies where the noise is lower.  Additional simulations for higher $Q$ are needed to address this question.

\subsection{Narrowband aLIGO}

The presence of a signal-recycling cavity in the aLIGO instruments will allow them to be tuned to have improved narrowband sensitivity at the expense of bandwith. Two parameters control the narrowband capabilities of the instruments~\cite{Meers:1988wp,Buonanno:2001cj,creighton:2011}: the signal recycling mirror transmissivity effectively sets the frequency bandwidth of the instrument, while the length of the signal recycling cavity (or equivalently the signal-recycling cavity tuning phase) controls the central frequency $f_R$ of the best sensitivity. By tuning one or more of the aLIGO detectors to operate in narrowband mode, it may be possible to improve estimates of the EOS parameters.

\begin{figure}[!htb]
\begin{center}
\includegraphics[width=80mm]{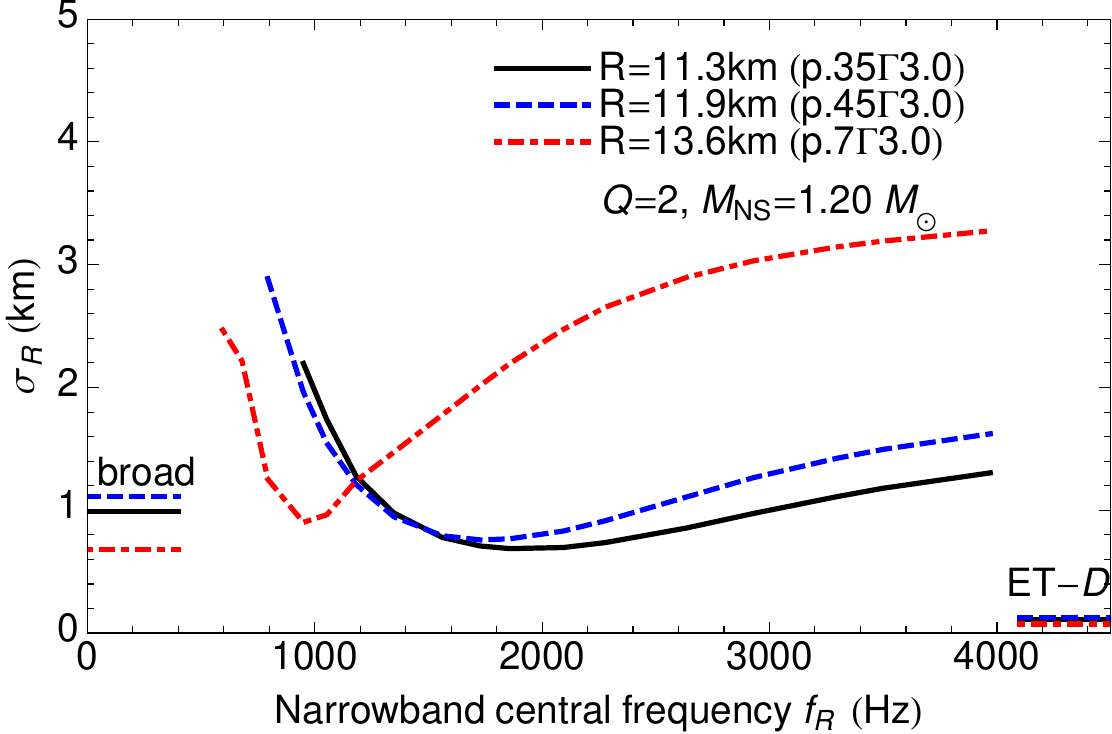}
\includegraphics[width=80mm]{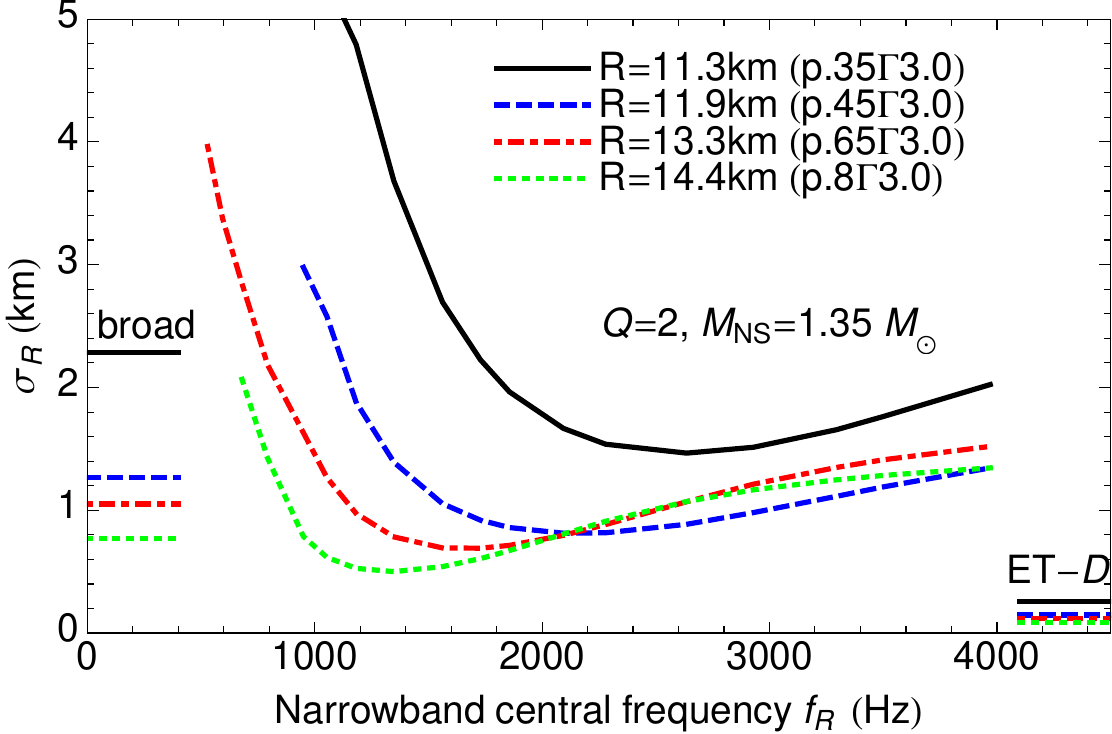}
\includegraphics[width=80mm]{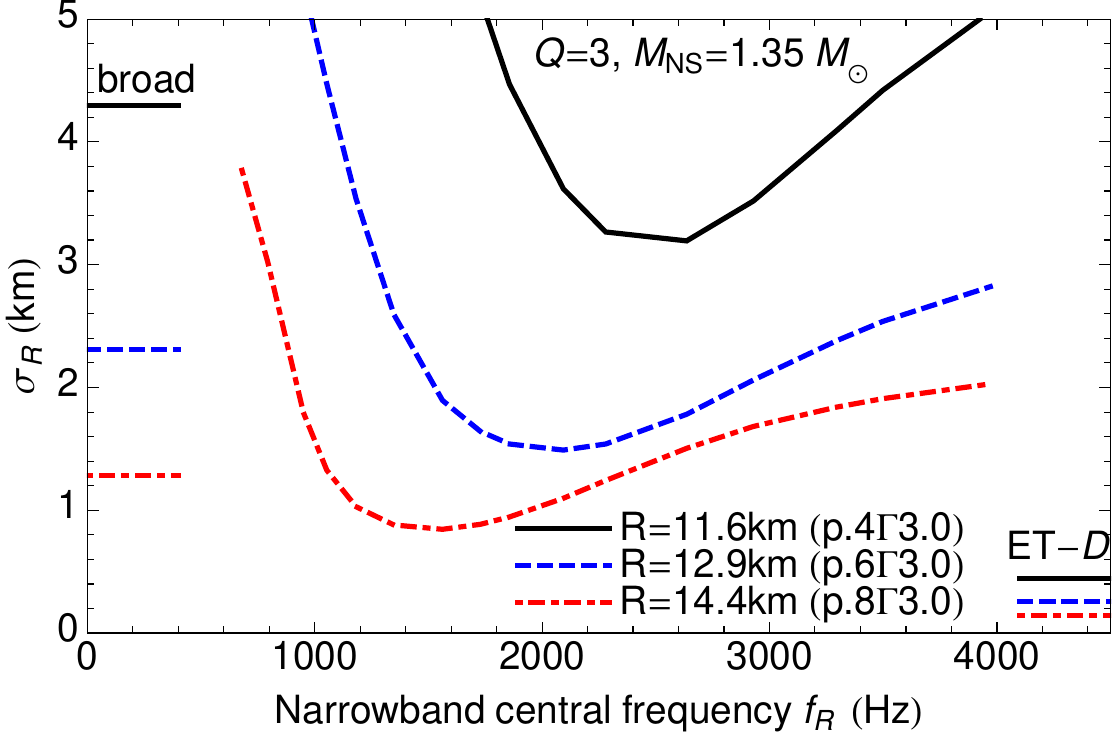}
\end{center}
\caption{ \label{fig:narrowbanderror}
1--$\sigma$ uncertainty in $R$ for different configurations of narrowband aLIGO and for different EOS.  $f_R$ defines the frequency where $S_n$ is a minimum as shown in Fig.~\ref{fig:noisecurves}.  Horizontal lines on the left and right indicate the corresponding 1--$\sigma$ errors for broadband aLIGO and ET-D respectively.  Matching and splicing conventions are those of Fig.~\ref{fig:waveform}.
}
\end{figure}

We have examined several narrowband tunings with central frequencies that vary between approximately $f_R=500$~Hz and $4000$~Hz.  These noise curves use a signal recycling mirror transmissivity of 0.011 and a signal-recycling cavity tuning phase ranging from $10^\circ$ down to $1^\circ$, and were generated using the program \texttt{gwinc}~\cite{gwinc}.  Three of these noise curves are shown in Fig.~\ref{fig:noisecurves}.  In Fig.~\ref{fig:narrowbanderror} we plot the 1--$\sigma$ uncertainty in NS radius $\sigma_R$ as a function of the narrowband central frequency $f_R$.  For the waveforms considered in this paper the optimal narrowband frequency is in the range $1000\textrm{ Hz} \lesssim f_R \lesssim 2500\textrm{ Hz}$ and depends on the mass ratio, NS mass, and EOS. Narrowband configurations usually give smaller errors than the broadband configuration if $f_R$ happens to be tuned to within a few hundred~Hz of the minimum for that BHNS event.  In Ref.~\cite{Hughes2002}, Hughes discussed a method for determining the best frequency $f_R$ to tune a narrowbanded detector to extract an EOS dependent cutoff frequency from a sequence of identical BNS inspirals. While this technique is not directly applicable to BHNS systems, which have different masses and spins, a similar approach could be used to combine multiple BHNS observations. 

\section{Discussion}
\label{sec:discussion}

\subsection{Results}

Using a large set of simulations incorporating a two-parameter EOS, we have found that the tidal deformability $\Lambda^{1/5}$, or equivalently the NS radius $R$, is the parameter that will be best extracted from BHNS waveforms.  These parameters can be estimated to 10--40\% with broadband aLIGO for an optimally oriented BHNS binary at 100~Mpc.  The narrowband aLIGO configuration can do slightly better if it is tuned to within a few hundred Hz of the ideal frequency for a given BHNS event.  The proposed Einstein Telescope will have an order-of-magnitude better sensitivity to the EOS parameters. 

Although we have used a particular EOS parametrization to show that $\Lambda$ is the parameter that is observed during BHNS coalescence, this result can be used to constrain any EOS model---an EOS based on fundamental nuclear theory in addition to a parametrized phenomenological EOS.  In particular, several parametrizations have recently been developed, including a spectral representation~\cite{Lindblom2010}, a reparametrization of the piecewise polytrope~\cite{OzelPsaltis2009}, and a generalization that also includes nuclear parameters~\cite{SteinerLattimerBrown2010}.

The results presented here can be compared with recent work to determine the mass and radius of individual NS in Type-1 X-ray bursts.  \"Ozel et al.~\cite{OzelBaymGuver2010} have obtained mass and radius measurements from several systems by simultaneously measuring the flux $F$, which is likely close to the Eddington value, and the blackbody temperature $T$ during X-ray bursts of systems with accurately determined distances.  During the burst, the emission area of the photosphere $F/(\sigma T^4)$ expands, contracts, then reaches a constant value, and \"Ozel et al.\ have argued that the final area corresponds to that of the NS surface.  They obtain estimates of NS mass and radii with $\mathcal{O}(10\%)$ 1--$\sigma$ uncertainty.  Steiner et al.~\cite{SteinerLattimerBrown2010} have also considered these systems, but argue that the final emission area does not necessarily correspond to that of the NS surface, and as a result obtain slightly smaller NS radii and larger uncertainties in the mass and radius.  These radius error estimates are slightly smaller than those for the BHNS systems we have considered at 100~Mpc.  However, we note that binary inspiral observations are subject to less systematic uncertainty due to questions of composition of the photosphere and associating it with the NS surface.  

The uncertainty in NS radius for the merger and ringdown of BHNS systems examined here is of roughly the same size as that found for the last few orbits up to merger of BNS systems at the same 100~Mpc distance~\cite{ReadMarkakisShibata2009, Read2011}.  BNS inspirals, however, will likely occur more frequently, and, including a tidally corrected inspiral--numerical hybrid, BNS systems are likely to have uncertainties that are smaller than BHNS systems by a factor of a few.  Considering the post-merger phase for BNS waveforms may also provide additional information.  Expected NS masses in both BNS and BHNS systems are slightly smaller than those measured for X-ray bursters which have accreted matter from their companion, so BNS and BHNS GW observations may complement X-ray burst observations by better constraining the lower density range of the EOS which is not well constained from X-ray burst observations~\cite{OzelPsaltis2009, OzelBaymGuver2010}.

\subsection{Future work}

We have used in this paper several simplifications and conventions which can significantly effect the accuracy to which EOS parameters can be extracted.  We list them below and describe how changing them would effect the parameter error estimates.

\begin{enumerate}

\item \textit{Finite length of numerical waveforms}\\
The BHNS waveforms used here include only the last $\sim$10 GW cycles of inspiral as well as the merger and ringdown, of which the first few cycles of inspiral are unreliable due to inexact initial data.  Matching the numerical waveform to a tidally corrected inspiral waveform instead of just the point-particle waveform will increase the overall departure from point-particle behavior by (i) creating a phase shift during the early inspiral, and more importantly (ii) adding to the phase of the late inspiral and merger the accumulated phase shift from the early inspiral -- a phase shift that is not already included in the stronger signal of the late inspiral. The tidal corrections are now known up to 1PN order.  For simulations with the current number of orbits, however, it appears that higher order tidal corrections will be needed to fully describe the late inspiral where matching to numerical waveforms is done.  We leave the issue of generating tidally corrected inspiral-numerical hybrid waveforms to future work.

\item \textit{Event rates}\\
Estimates of the detectability of EOS parameters in BNS systems are often given for an event at a distance of 100~Mpc, and we have used the same convention here to state the results above. The relevant event rate is, therefore, the expected number of detected events that will have an effective distance $D_{\mathrm{eff}} \leq 100\textrm{ Mpc}$. (The effective distance $D_{\mathrm{eff}}$ depends on the location of the binary and its inclination relative to the detector. For an optimally oriented and located binary, one finds $D=D_{\mathrm{eff}}$ while typically $D \leq D_{\mathrm{eff}}$.)
The aLIGO inspiral rates for BNS systems are highly uncertain with \{low, most likely, high\} estimates of \{0.01, 1, 10\}~Mpc$^{-3}$~Myr$^{-1}$~\cite{LIGORate2010} or \{0.004, 0.4, 4\}~yr$^{-1}$ with effective distance $D_{\mathrm{eff}} \leq 100\textrm{ Mpc}$.  Rates are even more uncertain for BHNS systems with rate estimates of \{0.0002, 0.01, 0.4\}~yr$^{-1}$ with effective distance $D_{\mathrm{eff}} \leq 100\textrm{ Mpc}$~\cite{LIGORate2010}. Since the uncertainty in EOS parameters scales linearly with distance [$\sigma_{\Lambda^{1/5}} = \sigma_{\Lambda^{1/5}, {100 \rm Mpc}} (D/100  \rm Mpc)$] and the event rate scales as $D^3$, the estimated detection rates of systems with effective distance $D_{\mathrm{eff}} \leq 400\textrm{ Mpc}$ are \{0.01, 1, 30\}~yr$^{-1}$ with a four-fold increase in uncertainty of $\Lambda^{1/5}$.  Fortunately, for $N_{\rm obs}$ identical events and $N_{\rm det}$ identical detectors, the uncertainty also scales as $\sigma_{\Lambda^{1/5}}/\sqrt{N_{\rm obs}N_{\rm det}}$.

\item \textit{Expected NS masses and mass ratios}\\
The simulations we used included realistic mass neutron stars of 1.2 and 1.35~$M_\odot$.  On the other hand, black hole masses are expected to be many times larger~\cite{BelczynskiKalogeraBulik2002}, with likely mass ratios closer to $Q\sim 7$ (for the canonical 10~$M_\odot$--1.4~$M_\odot$ system) than the $Q=2$ and 3 systems we examined here.  Additional simulations for mass ratios of 4 and 5 are in progress.

\item \textit{Spinning BH}\\
In this paper we have not examined the effect of a spinning BH.  The analytic results of Ref.~\cite{PannaraleRezzollaOhmeRead2011} indicated that spin does not significantly improve the sensitivity to $\Lambda$ for the inspiral up to the point of tidal disruption.  However, numerical simulations~\cite{Etienne2009, FoucartDuezKidderTeukolsky2011, KyutokuOkawaShibataTaniguchi2011} have shown that spin can strongly affect the dynamics near tidal disruption and the amount of matter left over in an accretion disk.  We have performed several tens of simulations of non-precessing BHNS systems with spinning BH with various BH spins, mass ratios, NS masses, and EOS parameters, and an analysis of how BH spin affects the detectability of EOS parameters will be the subject of the next paper. 

\item \textit{Correlations between parameters}\\
In our Fisher analysis we have assumed that the mass ratio, NS mass, and BH spin will be determined to sufficient accuracy during the inspiral to separate them from EOS effects during the merger and ringdown.  A full Fisher analysis using all of the BHNS parameters should be done to find the extent to which uncertainties in the other parameters alter measurability estimates of the EOS parameters.  

\end{enumerate}

Because BHNS waveforms smoothly deviate from corresponding BBH waveforms as $\Lambda$ increases, it is likely that one can find a good analytical approximation for the full inspiral, merger, and ringdown waveform by modifying analytical BBH waveforms.  Accurate waveforms for non-spinning BBH systems using the EOB approach have been developed~\cite{DamourNagar2009, PanBuonannoBoyle2011} and work to find EOB waveforms for spinning BBH systems is in progress~\cite{PanBuonannoBuchman2010, PanBuonannoFujita2011}.  Tidal interactions have also been incorporated into the EOB approach for BNS systems with good agreement with the inspiral waveform from numerical simulations when parametrized 2PN tidal interactions are fit to the numerical waveform~\cite{BaiottiDamour2010, BaiottiDamour2011}.  Another approach is to use phenomenological waveforms that fit the frequency domain post-Newtonian inspiral waveform to a phenomenological merger and ringdown for both spinning and non-spinning BBH systems~\cite{Santamaria2010}.  Both of these approaches may work for generating full analytic BHNS waveforms as well.  A complete description of the BHNS waveform will likely include corrections for the $l=3$ tidal field and other higher order corrections.  However, it is not clear given the current set of simulations that these effects would be observable with either aLIGO or a third generation detector such as ET.

\acknowledgments

We thank Alessandro Nagar for significant help with understanding the EOB formalism, Yi Pan and Alessandra Buonanno for providing EOB waveforms with spinning black holes, 
Jocelyn Read for providing routines used in the data analysis, and Jolien Creighton for generating narrowband noise curves.  This work was supported by NSF Grants PHY-1001515 and PHY-0923409, by Grant-in-Aid for Scientific Research (21340051), by Grant-in-Aid for Scientific Research on Innovative Area (20105004) of Japanese MEXT, and by a Grant-in-Aid of JSPS.  BL would also like to acknowledge support from a UWM Graduate School Fellowship and the Wisconsin Space Grant Consortium.  KK is also supported by the Grant-in-Aid for the Global COE Program ``The Next Generation of Physics, Spun from Universality and Emergence" of Japanese MEXT.''

\bibliography{nonspinning}

\appendix

\section{Numerically evaluating the Fisher matrix}
\label{app:numeval}

When an analytical representation of a waveform is not available, the partial derivatives in the Fisher matrix Eq.~(\ref{eq:fisher}) must be evaluated numerically.  There are several possible methods one can use, and we will examine their accuracy below.

\subsection{Finite differencing of $h(t;\theta)$}

The simplest method, and that used in Ref.~\cite{ReadMarkakisShibata2009}, is straightforward finite differencing of the signal $h = F_+h_+ + F_\times h_\times$.  For example, for five waveforms with values of an EOS parameter $\theta$ given by $\{\theta_{-2}, \theta_{-1}, \theta_{0}, \theta_{1}, \theta_{2}\}$ with equal spacing $\Delta \theta$, the three and five point central differences are given by 
\begin{eqnarray}
\frac{d h}{d \theta}
&=& \frac{\Delta_2 h}{\Delta \theta}+ \mathcal{O}((\Delta \theta)^2),\ \mbox{where}\nonumber\\
\frac{\Delta_2 h}{\Delta \theta} &:=&
\frac{-\frac{1}{2}h(t; \theta_{-1}) + \frac{1}{2}h(t; \theta_1)}{\Delta \theta}\\
\frac{d h}{d \theta}&=& 
\frac{\Delta_4 h}{\Delta \theta} + \mathcal{O}((\Delta \theta)^4),\ \mbox{where}
\nonumber\\
\frac{\Delta_4 h}{d \theta}&:=&
 \frac{\frac{1}{12}h(t;\theta_{-2}) - \frac{2}{3}h(t;\theta_{-1})  + \frac{2}{3}h(t;\theta_1) -\frac{1}{12}h(t;\theta_2)}{\Delta \theta}.\nonumber\\
\end{eqnarray}
This finite differencing method is useful when waveforms differ only slightly: at each time $t$, on the scale $\Delta\theta$ the function $h(t;\theta)$ is well approximated by the low order interpolating polynomials used to generate the finite differencing formulas.

This assumption fails when the waveforms used in the finite differencing are significantly out of phase with each other\footnote{The dephasing of numerical waveforms is even more significant for BNS inspiral.  We believe that Ref.~\cite{ReadMarkakisShibata2009} which used this method underestimated the derivatives in some cases by a factor of $\sim$2 or more, and thus overestimated the uncertainty in EOS parameters by the same factor.}.  The tidal interaction leads to a monotonically accumulating phase difference relative to a BBH waveform, implying that at a fixed time $t$ the function $h(\theta; t)$ is an oscillating function of $\theta$.  Now if an oscillating function $h[\Phi(\theta)]=\cos[\Phi(\theta)]$ has wavenumber $k=\Phi'(\theta)$ that varies slowly compared to $\Phi$, then $h'(\theta)$ is better approximated by $-\sin(\Phi)\Delta \Phi/\Delta \theta$ than by $\Delta\cos[\Phi(\theta)]/\Delta \theta$.  The assumption that $k$ is slowly varying is $k'\ll k^2$, $k'' \ll k^3$, and the error in, for example, each of the two second-order discretizations is given to order $\Delta\theta^2$ by
\begin{eqnarray}
\frac{d h}{d \theta} - \frac{\Delta_2 h}{\Delta \theta} &=& h(\theta) [\frac16 k^3 +O(kk', k'')]\Delta\theta^2,
\nonumber\\
\frac{d h}{d\theta} + \sin[\Phi(\theta)]\frac{\Delta_2 \Phi}{\Delta \theta} &=& h(\theta)\frac16 k''\Delta\theta^2,
\end{eqnarray}
with the error in the second expression much smaller than that in the first.  We consider two ways to take advantage of this difference in accuracy.

\subsection{Finite differencing of amplitude and phase}

The first is to decompose each complex waveform into an amplitude $A$ and accumulated phase $\Phi$
\begin{equation}
h_+(t;\theta) - i h_\times(t;\theta) = A(t;\theta) e^{-i \Phi(t;\theta)},
\end{equation}
where the accumulated phase of each waveform is a continuous function defined by $\Phi = -\arg(h_+ - ih_\times) \pm 2n\pi$ for some integer $n$, and at the starting time $t_i$ the accumulated phase of each waveform is chosen to be on the branch $n=0$.  The advantage of this method is that, at a fixed time, the functions $A(t;\theta)$ and  $\Phi(t;\theta)$ are non-oscillatory functions of $\theta$ even when the accumulated phase difference between two waveforms is significantly more than a radian.

With this decomposition the gravitiational wave signal is
\begin{equation}
h(t;\theta) = A(t;\theta)(F_+ \cos\Phi(t;\theta) + F_\times \sin\Phi(t;\theta)),
\end{equation}
and the derivative of $h$ is approximated by
\begin{eqnarray}
\frac{dh}{d\theta} &=& \frac{\Delta A}{\Delta \theta}(F_+ \cos\Phi + F_\times \sin\Phi) \nonumber \\
& & + A(-F_+ \sin\Phi + F_\times \cos\Phi)\frac{\Delta\Phi}{\Delta\theta}.
\end{eqnarray}
If an intermediate waveform is not available to provide the functions $A(t;\theta_0)$ and $\Phi(t;\theta_0)$, they can be evaluated by e.g.\ $A(t;\theta_0) = (A(t;\theta_{-1}) + A(t;\theta_{1})) / 2$.

We have found that this method works reasonably well for the inspiral waveform.  If, however, the amplitude of one of the numerical waveforms drops to zero, then the phase of the waveform becomes undefined.  Because the amplitude of the numerical BHNS waveforms fall to zero at different times for different EOS, as shown in Fig.~\ref{fig:hamp}, the derivative $d\Phi / d\theta$ becomes meaningless towards the end when the average amplitude is still nonzero.  It is likely one could work around this difficulty.  However, we choose instead to use another more robust method.

\subsection{Finite differencing of Fourier transform}

Because we will need to calculate the Fourier transform of the derivative $dh / d\theta$ to find the Fisher matrix, we first Fourier transform each waveform and then evaluate the numerical derivative.  Since the derivative $d / d\theta$ commutes with the Fourier transform, the Fisher matrix can be written explicitly as
\begin{equation}
\left(\frac{\partial h}{\partial \theta^A}\left|\frac{\partial h}{\partial \theta^B}\right.\right)
=4{\rm Re} \int_{f_i}^{f_f} \frac{\frac{\partial\tilde h}{\partial \theta^A} \frac{\partial\tilde h^*}{\partial \theta^B}}{S_n(f)}\,df,
\end{equation}
where the contribution to the integral below $f_i$ and above $f_f$ is negligable.

As in the second method we break up each Fourier transformed waveform into amplitude $A(f;\theta)$ and accumulated phase $\Phi(f;\theta)$
\begin{equation}
\tilde h(f;\theta) = A(f;\theta) e^{- i \Phi(f;\theta)},
\end{equation}
where the phase of each waveform at $f_i$ is on the $n=0$ branch cut.  As demonstrated by Figs.~\ref{fig:spectrum} and~\ref{fig:phasefreq}, both the amplitude and phase are non-oscillatory functions of $\theta$ at a fixed frequency $f$, and can be well approximated by a low-order polynomial.  In contrast to the accumulated phase of the complex numerical waveform $h_+ - ih_\times$, the accumulated phase of the Fourier transform of the strain $\tilde h$ is always well defined for numerical BHNS waveforms in the frequency range $f_i$ to $f_f$.

Finally, we find that one obtains better accuracy by differentiating $\ln A$ instead of $A$, 
decomposing $\tilde h$ as
\begin{equation}
\label{eq:logampphase}
\tilde h(f;\theta) = e^{\ln A(f;\theta) - i \Phi(f;\theta)}.
\end{equation}
The derivative is now approximated by
\begin{equation}
\frac{\partial \tilde h}{\partial\theta} 
= e^{\ln A - i \Phi}\left( \frac{\Delta\ln A}{\Delta\theta} - i \frac{\Delta \Phi}{\Delta\theta} \right).
\end{equation}
interpolating when needed to evaluate $\ln A$ and $\Phi$ at the midpoint.

\subsection{Parameter spacing}

Finally, we note that the EOS parameter spacing must be carefully chosen.  If two waveforms are too close in parameter space, the error in each waveform will dominate over the truncation error due to finite differencing.  The most significant source of this error comes from the spurius oscillations in the amplitude of the Fourier transform in the frequency range $\sim 500$--800~Hz (see Fig.~\ref{fig:spectrum}) that result from joining the EOB and numerical waveforms which are not exactly the same in the matching window.  We find that the integrand of the Fisher matrix is often erratic in the range $\sim 500$--800~Hz when using the smallest parameter spacing available.  However, when the spacing is increased, the integrand is smoother in this frequency range and its contribution to the integral is significantly reduced.  For the mass ratio $Q=2$, we find that a spacing between waveforms of $\Delta\log(p_1 / ({\rm dyne\ cm}^{-2})) = 0.1$ for the first EOS parameter is often sufficiently large to reduce this problem, while a spacing of $\Delta\Gamma = 0.6$ for the second EOS parameter is the minimum spacing one can use.  For $Q=3$, we have found that a spacing of $\Delta\log(p_1 / ({\rm dyne\ cm}^{-2})) \ge 0.2$ is necessary to reduce this problem.  

In addition, if the EOS parameters of two waveforms lie near the same degenerate contour where waveforms are identical (e.g.\ a contour of constant $\Lambda$ which can be nearly identical to a line of constant $\Gamma$), the error in each waveform will again dominate the truncation error even if the EOS parameters are widely spaced.  For our two-dimensional EOS parameter space, this problem can be solved by transforming the parameter space such that points that originally formed a $\times$ pattern now form a $+$ pattern, and in the transformed parameter space the new axes are not alligned with a degenerate contour.  The Fisher matrix can be calculated in the transformed parameter space then transformed back to the original parameter space.  

We find that as long as these two requirements are met, the uncertainties in $\sigma_{\lambda^{1/5}}$ and $\sigma_{R}$ have only an $\mathcal{O}(20\%)$ fractional dependence on the EOS parameter spacing.  However, the dependence of the orientation of the error ellipses on the EOS parameter spacing does not allow one to distinguish between $\Lambda$ and $R$ as the best extracted parameter.

\section{Prescription for calculating EOB waveforms}
\label{app:eob}

In this appendix we compile the necessary ingredients needed to produce nonspinning BBH waveforms using the effective one body (EOB) formalism first introduced in Ref.~\cite{BuonannoDamour1999}.  The version used here is exactly that of Ref.~\cite{DamourNagar2009}, and is described in more detail in a review~\cite{DamourNagar2009review}.  The only ingredients not listed here are terms for the re-sumed waveform in Ref.~\cite{DamourIyerNagar2009} and coefficients to determine the ringdown waveform found in Ref.~\cite{BertiCardosoWill2006}.

\subsection{Hamiltonian dynamics}

In the EOB formalism the two-body dynamics are replaced by a test particle of reduced mass $\mu=m_1m_2/M$ moving in a modified Schwarzschild metric of total mass $M=m_1+m_2$ given by\footnote{The expressions below will be written exclusively in terms of rescaled dimensionless quantities.  The coordinates $(T, R, \phi)$ and conjugate momenta $(P_R, P_\phi)$ have been rescaled to dimensionless coordinates $(t, r, \phi)$ and momenta $(p_r, p_\phi)$ given by: $t=T/M$ and $r=R/M$ for the coordinates, and $p_r=P_R/\mu$, $p_\phi = P_\phi/\mu M$ for the conjugate momenta.  Other quantities are then rescaled in the following way: $\omega = M\Omega = M d\phi/dT$ is the angular velocity, $\hat D = D/M$ is the distance to the source, $\hat{H} = H/\mu$ and $\hat{H}_{\rm eff} = H_{\rm eff}/\mu$ are the Hamiltonian and effective Hamiltonian, and $\mathcal{\hat{F}}_\phi = \mathcal{F}_\phi/\mu$ is the radiation reaction force.}
\begin{equation}
ds^2 = -A(r)dt^2 + B(r)dr^2 + r^2(d\theta^2 + \sin^2\theta d\phi^2).
\end{equation}
The metric potentials $A$ and $B$ can be calculated from post-Newtonian theory.  The first function is
\begin{equation}
\label{eq:Aofu}
A(u) = P^1_5[1 - 2u + 2\nu u^3 + \left(\frac{94}{3}-\frac{41\pi^2}{32}\right)\nu u^4 + a_5\nu u^5 + a_6\nu u^6],
\end{equation}
where $u=1/r$, $\nu=\mu/M$ is the symmetric mass ratio, and $P^m_n[\cdot]$ denotes a Pad\'{e} approximant of order $m$ in the numerator and $n$ in the denominator.  The 4 and 5 PN coefficients, $a_5$ and $a_6$, are fit to numerical BBH waveforms.  The values that give the optimal fit form a degenerate curve in the $a_5$--$a_6$ parameter space, and the specific values chosen here are $(a_5, a_6) = (0, -20)$.  The second potential is rewritten as
\begin{equation}
D(r) = B(r)A(r),
\end{equation}
and has been calculated to 2PN order
\begin{equation}
D(u) = P^0_3[1-6\nu u^2 + 2(3\nu-26)\nu u^3].
\end{equation}

The motion of the EOB particle of mass $\mu$ is determined by the Hamiltonian
\begin{equation}
\hat{H} = \frac{1}{\nu}\sqrt{1+2\nu(\hat{H}_{\rm eff}-1)},
\end{equation}
where
\begin{equation}
\hat{H}_{\rm eff} = \sqrt{A(1/r)\left(1+\frac{p_\phi^2}{r^2}+\frac{p_r^2}{B} + 2\nu(4-3\nu)\frac{p_r^4}{r^2}\right)}
\end{equation}
is the effective Hamiltonian.  The equations of motion given this conservative Hamiltonian $\hat{H}$ and a dissipative radiation-reaction force $\mathcal{\hat{F}}_i$ are
\begin{eqnarray}
\frac{dr}{dt} &=& \frac{\partial\hat{H}}{\partial p_r} \\
\frac{d\phi}{dt} &=& \frac{\partial\hat{H}}{\partial p_\phi} = \omega \\
\frac{dp_r}{dt} &=& -\frac{\partial\hat{H}}{\partial r} +\mathcal{\hat{F}}_r\\
\frac{dp_\phi}{dt} &=& -\frac{\partial\hat{H}}{\partial \phi} +\mathcal{\hat{F}}_\phi.
\end{eqnarray}
Here, $\frac{\partial\hat{H}}{\partial \phi} = 0$ because the EOB Hamiltonian does not have an explicit $\phi$ dependence.  In addition, for circularized binary inspiral the radial component of the radiation-reaction force $\mathcal{\hat{F}}_r$ is of higher post-Newtonian order than the tangential component, so it is set to zero.

To increase resolution near the black hole, the radial coordinate can be rewritten in terms of a tortoise coordinate~\cite{DamourNagar2007} defined by
\begin{equation}
\frac{dr_*}{dr} = \left(\frac{B}{A}\right)^{1/2}.
\end{equation}
The new radial momentum is then $p_{r_*} = (A/B)^{1/2}p_r$.  Using this definition, the effective Hamiltonian becomes
\begin{equation}
\hat{H}_{\rm eff} = \sqrt{p_{r_*}^2+A(1/r)\left(1+\frac{p_\phi^2}{r^2}+2\nu(4-3\nu)\frac{p_{r_*}^4}{r^2}\right)}
\end{equation}
where the parts that are 4PN and higher are neglected.  (The 4 and 5 PN terms are however accounted for in the free parameters $a_5$ and $a_6$ which were fit to numerical waveforms).  The equations of motion become
\begin{eqnarray}
\frac{dr}{dt} &=& \frac{A}{\sqrt{D}} \frac{\partial\hat{H}}{\partial p_{r_*}} \label{eq:eomr}\\
\frac{d\phi}{dt} &=& \frac{\partial\hat{H}}{\partial p_\phi} = \omega \label{eq:eomphi}\\
\frac{dp_{r_*}}{dt} &=& -\frac{A}{\sqrt{D}} \frac{\partial\hat{H}}{\partial r} \label{eq:eompr}\\
\frac{dp_\phi}{dt} &=& \mathcal{\hat{F}}_\phi. \label{eq:eompphi}
\end{eqnarray}

\subsection{Radiation reaction}

For the radiation reaction term $\mathcal{\hat F}_\phi$, which is written in terms of the PN parameter $x$, we will need a way to write $x$ in terms of the dynamical variables.  The usual method is to use the Newtonion potential $1/r$ and velocity squared $(\omega r)^2$ as PN counting parameters and then rewrite them in terms of the gauge invariant angular velocity $\omega$ using the Kepler law $\omega^2 r^3 = 1$ which holds in the Newtonian limit, and for circular orbits, in the Schwarzschild ($\nu \rightarrow 0$) limit.  The Kepler relation can be extended to circular orbits in the EOB metric by defining a new radial parameter, $r_\omega = r \psi^{1/3}$, where
\begin{equation}
\psi(r, p_\phi) = \frac{2}{r^2}\left(\frac{dA}{dr}\right)^{-1}\left[1+2\nu\left(\sqrt{A(r)\left(1+\frac{p_\phi^2}{r^2}\right)}-1\right)\right],
\end{equation}
for which $\omega^2 r_\omega^3 = 1$ holds for all circular orbits.  In addition, for noncircular orbits (in particular for the plunge), this relation also relaxes the quasicircular condition by not requiring that the Kepler relation hold.  The specific choice of PN parameter used here is
\begin{equation}
x = (\omega r_\omega)^2.
\end{equation}
See Ref.~\cite{DamourGopakumar2006} for an extensive discussion.

The radiation reaction term $\mathcal{\hat{F}}_\phi$ used in Ref.~\cite{DamourNagar2009} takes the form of a summation over all multipoles
\begin{equation}
\mathcal{\hat{F}}_\phi = -\frac{1}{8 \pi \nu \omega}{\sum_{\ell=2}^{8}\sum_{m=1}^{\ell}} (m\omega)^2 |\hat D h_{\ell m}|^2.
\end{equation}
Instead of the standard Taylor expanded version of $h_{\ell m}$ which can be found in Ref.~\cite{Kidder2008}, Ref.~\cite{DamourIyerNagar2009} decomposed the waveform into a product of terms:
\begin{equation}
\label{eq:h22}
h_{22} = h_{22}^{\rm Newt} \hat{S}_{\rm eff} T_{22} e^{i \delta_{22}} f_{22}(x) f_{22}^{\rm NQC}
\end{equation}
for $\ell=m=2$, and
\begin{equation}
\label{eq:hlm}
h_{\ell m} = h_{\ell m}^{\rm Newt} \hat{S}_{\rm eff} T_{\ell m} e^{i \delta_{\ell m}} \rho_{\ell m}^\ell(x)
\end{equation}
for the other values of $\ell$ and $m$.  The leading Newtonian part $h_{\ell m}^{\rm Newt}$ is given in the usual form as a function of $x$ 
\begin{equation}
h_{\ell m}^{\rm Newt} = \frac{\nu}{\hat D} n_{\ell m} c_{\ell+\epsilon}(\nu) x^{(\ell+\epsilon)/2} Y^{\ell-\epsilon, -m}\left(\frac{\pi}{2}, \phi\right)
\end{equation}
where the coefficients $n_{\ell m}$ and $c_{\ell+\epsilon}(\nu)$ are defined by Eqs.~(5--7) of Ref.~\cite{DamourIyerNagar2009}, and the parity $\epsilon$ is 0 for $
\ell+m$ even and 1 for $\ell+m$ odd.

The PN terms in the resummation which had been written as functions of $x$ in Ref.~\cite{DamourIyerNagar2009} are now written in terms of the dynamical variables.  The effective source term $\hat{S}_{\rm eff}$ becomes~\cite{NagarPC}
\begin{equation}
\hat{S}_{\rm eff}(r, p_{r_*}, p_\phi) = 
\left\{\begin{array}{ll}
\hat{H}_{\rm eff}(r, p_{r_*}, p_\phi) & \epsilon = 0\\
\frac{p_\phi}{r_\omega^2\omega} & \epsilon = 1
\end{array}\right..
\end{equation}
The tail term is
\begin{equation}
T_{\ell m}(r, p_{r_*}, p_\phi) = \frac{\Gamma(\ell + 1 - 2i\hat{\hat{k}})}{\Gamma(\ell + 1)} e^{\pi\hat{\hat{k}}} e^{2i\hat{\hat{k}}\ln{2kr_0}},
\end{equation}
where $\hat{\hat{k}} = \nu m \hat H(r, p_{r_*}, p_\phi) \omega (r, p_{r_*}, p_\phi)$, $k = m\omega (r, p_{r_*}, p_\phi)$, and $r_0 = 2$.  The phase of this tail term is corrected with a term of the form $e^{i\delta_{\ell m}}$.  The first ten $\delta_{\ell m}$ are given in Eqs.~(20--29) of Ref.~\cite{DamourIyerNagar2009}.  The first one is
\begin{equation}
\delta_{22} = \frac{7}{3}y^{3/2} + \frac{428\pi}{105}y^3 - 24\nu\bar{y}^{5/2},
\end{equation}
where $y = (\nu \hat H(r, p_{r_*}, p_\phi) \omega (r, p_{r_*}, p_\phi))^{2/3}$ and $\bar y$, which has several possible forms, is chosen to be $\bar y = \omega^{2/3}$~\cite{NagarPC}.  Finally, the remainder term of the resummation $f_{\ell m}$ is expanded in powers of $x$.  For $\ell = m = 2$ this is then re-summed with a Pad\'{e} approximant
\begin{equation}
f_{22}(x) = P_2^3[f_{22}^{\rm Taylor}(x)],
\end{equation}
where
\begin{widetext}
\begin{eqnarray}
f_{22}^{\rm Taylor}(\nu, x) &=& 1 + \frac{55\nu-86}{42}x + \frac{2047\nu^2 - 6745\nu - 4288}{1512}x^2 \nonumber\\
& & + \left(\frac{114635\nu^3}{99792}-\frac{227875\nu^2}{33264} + \frac{41\pi^2\nu}{96} - \frac{34625\nu}{3696} - \frac{856}{105}\rm{eulerln}_2(x) + \frac{21428357}{727650}\right)x^3 \nonumber\\
& & + \left(\frac{36808}{2205}\rm{eulerln}_2(x) - \frac{5391582359}{198648450}\right)x^4
+ \left(\frac{458816}{19845}\rm{eulerln}_2(x) - \frac{93684531406}{893918025}\right)x^5,
\end{eqnarray}
and the $\rm{eulerln}_m(x) = \gamma_E + \ln 2 + \frac{1}{2} \ln x + \ln m$ terms are treated as coefficients when calculating the Pad\'{e} approximant.  For the other values of $\ell$ and $m$, $f_{\ell m}$ is re-summed in the form $f_{\ell m} = \rho_{\ell m}^\ell$.  The quantity $\rho_{\ell m}$ is given in Eqs.~(C1--C35) of Ref.~\cite{DamourIyerNagar2009}.  $\rho_{21}$ is for example
\begin{eqnarray}
\rho_{21} &=& 1 + \left(\frac{23\nu}{84}-\frac{59}{56}\right)x + \left(\frac{617\nu^2}{4704} - \frac{10993\nu}{14112} -\frac{47009}{56448}\right)x^2 + \left(\frac{7613184941}{2607897600}-\frac{107}{105}\rm{eulerln}_1(x)\right)x^3 \nonumber\\
& & + \left(\frac{6313}{5880}\rm{eulerln}_1(x) - \frac{1168617473883}{911303737344}\right)x^4.
\end{eqnarray}
\end{widetext}

The final product in the resummation of $h_{22}$ is a next to quasicircular (NQC) correction term that is used to correct the dynamics and waveform amplitude during the plunge 
\begin{equation}
f_{22}^{\rm NQC}(a_1, a_2) = 1 + \frac{a_1 p_{r_*}^2}{(r\omega)^2} + \frac{a_2 \ddot{r}}{r\omega^2}.
\end{equation}
The free parameters $a_1$ and $a_2$ are determined by the following conditions: (i) the time when the orbital frequency $\omega$ is a maximum (the EOB merger time $t_M$) coincides with the time when the amplitude $|h_{22}|$ is a maximum, and (ii) the value of the maximum amplitude is equal to a fitting function that was fit to several BBH simulations and is given by
\begin{equation}
|h_{22}|_{\rm max}(\nu) = 1.575\nu(1-0.131(1-4\nu)).
\end{equation}

\subsection{Integrating the equations of motion}

The equations of motion are solved by starting with initial conditions $\{r_0, \phi_0, p_{r_*0}, p_{\phi 0}\}$ and numerically integrating the equations of motion.  In this paper we are interested in long, zero-eccentricity orbits.  This can be achieved in the EOB framework by starting the integration with large $r$, where radiation reaction effects are small, and using the quasicircular condition $p_{r_*} = 0$.  Eq.~(\ref{eq:eompr}) then becomes
\begin{equation}
\frac{\partial H}{\partial r}(r, p_{r_*}=0, p_\phi) = 0
\end{equation}
and results in the condition
\begin{equation}
p_\phi^2 = - \frac{\frac{d}{du}A(u)}{\frac{d}{du}(u^2A(u))}
\end{equation}
for $p_\phi$.  If this quasicircular initial condition is used for smaller $r$, the radiation reaction term is no longer negligable, and this initial condition will result in eccentric orbits.  If desired, one can use an initial condition that more accurately approximates a zero eccentricity inspiral such as post-circular or post-post-circular initial conditions with nonzero $p_{r_*}$~\cite{DamourNagarDorband2008}.

To numerically solve Eqs.~(\ref{eq:eomr}--\ref{eq:eompphi}), they must be written as a system of first order equations.  However, the term $\mathcal{\hat{F}}_\phi$ in Eq.~(\ref{eq:eompphi}) contains the square of $\ddot{r}$ from the NQC term $f_{22}^{\rm NQC}$ in $h_{22}$.  Since $f_{22}^{\rm NQC}$ gives a small correction of order 10\% during the plunge, the easiest method, and that used in Ref.~\cite{DamourNagar2009}, is iteration~\cite{NagarPC}:  (i) First solve the system of equations with $f_{22}^{\rm NQC}$ set to one.  (ii) Use the solution of Eqs.~(\ref{eq:eomr}--\ref{eq:eompphi}) to evaluate $\ddot{r}$ and the other quantities in $f_{22}^{\rm NQC}$.  (iii) Re-solve the equations of motion with the NQC coefficients no longer set to one.  (iv) Repeat steps (ii) and (iii) until the solution converges to the desired accuracy.  In practice this iteration only needs to be done roughly 2--5 times.

A second method is to directly rewrite Eq.~(\ref{eq:eompphi}) as a first order equation.  This can be done by replacing $\ddot{r}$ in the NQC term on the right hand side with an expression containing $\dot{p}_\phi$ and then solving for $\dot{p}_\phi$.  The equations of motion~(\ref{eq:eomr}--\ref{eq:eompphi}) and the chain rule give

\begin{eqnarray}
\ddot{r} &=& \frac{d}{dt}\left( \frac{A}{\sqrt{D}} \frac{\partial\hat{H}}{\partial p_{r_*}} \right)\\
&=& L + M + N\dot{p}_\phi
\end{eqnarray}
where
\begin{eqnarray}
L &=& \frac{1}{2} \frac{\partial}{\partial r}\left[ \frac{A^2}{D}\left(\frac{\partial\hat{H}}{\partial p_{r_*}}\right)^2 \right]\\
M &=& -\frac{A^2}{D} \frac{\partial\hat{H}}{\partial r} \frac{\partial^2\hat{H}}{\partial p_{r_*}^2}\\
N &=& \frac{A}{\sqrt{D}} \frac{\partial^2\hat{H}}{\partial p_{r_*} \partial p_\phi}.
\end{eqnarray}
Plugging this expression into Eq.~(\ref{eq:eompphi}) yields an equation quadratic in $\dot{p}_\phi$ which can be solved exactly if desired.  To first order in the NQC correction term, Eq.~(\ref{eq:eompphi}) now becomes the first order equation 
\begin{equation}
\frac{dp_\phi}{dt} = \frac{\mathcal{\hat{F}}_{\phi, \rm Higher} + \mathcal{\hat{F}}_{\phi, 22}^{\rm QC}\left[ 1 +  
2\frac{a_1 p_{r_*}^2}{(r\omega)^2} + 2\frac{a_2}{r\omega^2}(L+M) \right]}
{1 - 2\mathcal{\hat{F}}_{\phi, 22}^{\rm QC} \frac{a_2}{r\omega^2}N},
\end{equation}
where
\begin{equation}
\mathcal{\hat{F}}_{\phi, \rm Higher} = -\frac{1}{8 \pi \nu \omega}\underset{(\ell, m)\ne(2, 2)}{\sum_{\ell=2}^{8}\sum_{m=1}^{\ell}} (m\omega)^2 |\hat D h_{\ell m}|^2
\end{equation}
includes just the higher order terms $(\ell, m)\ne(2, 2)$, and
\begin{equation}
\mathcal{\hat{F}}_{\phi, 22}^{\rm QC} = -\frac{1}{8 \pi \nu \omega} (2\omega)^2 |\hat D h_{22}^{\rm QC}|^2.
\end{equation}
The QC in $h_{22}^{\rm QC}$ means that the NQC term $f_{22}^{\rm NQC}$ has been factored out.

The solution to the equations of motion $\{r(t), \phi(t), p_{r_*}(t), p_\phi(t)\}$ are then plugged back into Eqs.~(\ref{eq:h22}--\ref{eq:hlm}) to give the waveform $h_{\ell m}^{\rm inspiral}(t)$.

\subsection{Ringdown}

In the EOB formalism the ringdown waveform of the final Kerr black hole is smoothly matched onto the inspiral waveform at the EOB merger time $t_M$.  The mass of the black hole remnant is given by the energy of the EOB particle at the merger time $t_M$
\begin{equation}
M_{\rm BH} \equiv \mu \hat{H}(t_M) = M \sqrt{1+2\nu(\hat{H}_{\rm eff}(t_M)-1)},
\end{equation}
and the Kerr parameter is given by the final angular momentum of the EOB particle~\cite{BuonannoDamour2000}
\begin{equation}
\hat{a}_{\rm BH} \equiv \frac{P_\phi(t_M)}{M_{\rm BH}^2} = \frac{\nu p_\phi(t_M)}{1 + 2\nu(\hat{H}_{\rm eff}(t_M)-1)}.
\end{equation}

The ringdown waveform is given by the first five positive quasinormal modes (QNM) for a black hole of mass $M_{\rm BH}$ and spin $\hat a_{\rm BH}$:
\begin{equation}
h_{22}^{\rm ringdown}(t) = \frac{1}{\hat D} \sum_{n=0}^{4} C_{22n}^+ e^{-\sigma_{22n}^+ (t-t_M)},
\end{equation}
where $\sigma_{22n}^+ = \alpha_{22n}+i\omega_{22n}$ is the nth complex $\ell=m=2$ QNM frequency for a Kerr BH with mass $\hat{M}_{\rm BH}$ and spin $\hat{a}_{\rm BH}$, and $C_{22n}^+$ are complex constants that determine the magnitude and phase of each QNM.  The amplitude of the negative frequency modes is small~\cite{DamourNagar2007}.  The first three QNMs have been tabulated in Ref.~\cite{BertiCardosoWill2006}, and fitting formuli are also provided.  The QNM frequency $\omega_{22n}$ can be approximated by
\begin{equation}
M_{\rm BH} \omega_{22n} = f_1 + f_2(1 - \hat a_{\rm BH})^{f_3},
\end{equation}
and the inverse damping time $\alpha_{22n}$ is given in terms of the quality factor approximated by
\begin{equation}
\frac{1}{2} \frac{\omega_{22n}}{\alpha_{22n}} = q_1 + q_2(1 - \hat a_{\rm BH})^{q_3}.
\end{equation}
The coefficients for $n=0$--2 can be found in table VIII of Ref.~\cite{BertiCardosoWill2006}.  For $n = 3$--4, $\alpha_{22n}$ and $\omega_{22n}$ can be linearly extrapolated from the values for $n=1$ and 2 as was done in Ref.~\cite{DamourNagarDorband2008}.

The constants $C_{22n}^+$ are determined by requiring that the  inspiral and ringdown waveforms be continuous on a ``matching comb'' centered on the EOB merger time $t_M$.  Specifically, at the times $\{t_M-2\delta, t_M-\delta, t_M, t_M+\delta, t_M+2\delta\}$ we require $h_{22}^{\rm inspiral}(t) = h_{22}^{\rm ringdown}(t)$.  In Ref.~\cite{DamourNagar2009}, $\delta$ was chosen to be equal to $2.3M_{\rm BH}/M$.  This gives 5 complex equations for the 5 unknown complex coefficients $C_{22n}^+$.  

The full inspiral plus ringdown waveform is then given by
\begin{equation}
h_{22}(t) =\
\left\{\begin{array}{lc}
h_{22}^{\rm inspiral}(t) & \, t<t_M \\
h_{22}^{\rm ringdown}(t) & \, t>t_M
\end{array}\right..
\end{equation}

\end{document}